\documentclass[a4paper,11pt]{article}
\usepackage{jheppub}
\usepackage[T1]{fontenc} 
\usepackage{amssymb,amsmath}
\usepackage[caption=false]{subfig}

\usepackage{breqn}
\usepackage{bbm}

\newcommand\eps{\epsilon}
\newcommand\cN{  {\cal N}  }

\newcommand{\bea}{\begin{eqnarray}}
\newcommand{\eea}{\end{eqnarray}}

\newcommand\bu{{\beta}_u}
\newcommand\bv{{\beta}_v}
\newcommand\buv{{\beta}_{uv}}

\renewcommand\v{v}

\newcommand\Li{{\rm Li}}

\def\GG#1#2#3{G_{{#1},{#2},{#3}}}  
  
\def\gYM{g_{{\rm YM}}}
\def\be{\begin{equation}}
\def\ee{\end{equation}}

\DeclareMathOperator\IM{Im}

\renewcommand{\d}{{\mathrm{d}}}

\newcommand{\mi}[1]{\vec{ #1 }} 

\allowdisplaybreaks[4]

\usepackage{tikz}
\usetikzlibrary{positioning,arrows}
\usepgflibrary{arrows.meta}
\usetikzlibrary{decorations.pathmorphing}
\usetikzlibrary{decorations.markings}
\usetikzlibrary{calc}

\tikzset{
Wilson/.style={double,postaction={decorate}, decoration={markings,mark=at position .1 with {\arrow{Stealth[scale=1]}},mark=at position .98 with {\arrow{Stealth[scale=1]}}}},
Wilson_blank/.style={double},
Wilson_arrow/.style={double,postaction={decorate}, decoration={markings,mark=at position .7 with {\arrow{Stealth[scale=1.1]}}}},
Scalar/.style={dashed},
Gluon/.style={decorate, draw=black, decoration={coil,aspect=0.5, post length = 2pt, pre length = 2pt, segment length=3pt,amplitude=3pt}},
Incoming/.style={dashed, postaction={decorate}, decoration={markings,mark=at position 0.6 with {\arrow{Stealth[scale=1.4,reversed]}}}}
 }

\tikzset{
    halfarrow/.style={postaction={decorate},
        decoration={markings,mark=at position .5 with
       {\arrow{Stealth[scale=1.2]}}}}}
       
\tikzset{axes_style/.style={->}} 
\tikzset{path_style/.style={halfarrow,blue}} 
\tikzset{help_lines_style/.style={dashed}} 
\tikzset{ana_strct_style/.style={dotted}} 
\tikzset{tick_style/.style={}} 
\tikzset{pointer_style/.style={>={Stealth}}} 

\preprint{MITP/18-004} 
\title{
Subleading Regge limit from a soft anomalous dimension
}

\author[a]{Robin Br\"user,}
\author[b]{Simon Caron-Huot,}
\author[a,c]{Johannes M.\ Henn}
\affiliation[a]{PRISMA Cluster of Excellence, Mainz University, Germany}
\affiliation[b]{McGill University, Montr\'eal, Canada}
\affiliation[c]{KITP, Santa Barbara, USA}
\emailAdd {brueser@uni-mainz.de}
\emailAdd {schuot@physics.mcgill.ca}
\emailAdd {henn@uni-mainz.de}

\abstract{
Wilson lines capture important features of scattering amplitudes,
for example soft effects relevant for infrared divergences, and the Regge limit.
Beyond the leading power approximation, corrections to the eikonal picture have to be taken into account. 
In this paper, we study such corrections in a model of massive 
scattering amplitudes in ${\mathcal N}=4$ super Yang-Mills, in the planar limit, where
the mass is generated through a Higgs mechanism.
Using known three-loop analytic expressions for the scattering amplitude,
we find that the first power suppressed term has a very simple form, equal to a single power law.
We propose that its exponent is governed by the anomalous
dimension of a Wilson loop with a scalar inserted at the cusp, and we provide perturbative evidence for this proposal.
We also analyze other limits of the amplitude and conjecture
an exact formula for a total cross-section at high energies.}


\begin{document}

\maketitle
\flushbottom

\section{Introduction}  

Future physics analyses at the LHC will require conceptional advances in the theoretical
understanding of scattering processes. 
One new frontier will be higher-loop processes depending on many mass and
kinematic scales, e.g. when considering mixed QCD and electroweak processes.
While in some cases a numerical approach may be feasible and adequate,
it seems clear that conceptual breakthroughs will be driven by new analytic ideas. 

When dealing with processes depending on many scales, an
important question is to understand in which situations systematic
expansions can be applicable, and how the latter can be obtained systematically.
One particularly interesting and important limit is the eikonal limit,
which describes emission of soft radiation. At leading power, the 
physics is described by correlation functions of Wilson lines.
Many recent papers are dedicated to studying power corrections
to the eikonal limit \cite{Laenen:2010uz,Luna:2016idw,Moult:2017rpl}.

It is often extremely helpful to have a toy model at hand for developing new
ideas. Among various Yang-Mills theories, the $\mathcal{N}=4$ super Yang-Mills theory stands out
as a particularly nice model. It is often dubbed the hydrogen atom of quantum field theory,
due to a hidden symmetry that is the generalization of the Laplace-Runge-Lenz symmetry
of the hydrogen atom. 
In order to be able to study massive scattering amplitudes,
we give a vacuum expectation value to some of the scalar fields in the model.
In this way, we can consider four-particle amplitudes depending on two Mandelstam
variables and a mass. The amplitudes are ultraviolet and infrared finite, so that they 
can be evaluated directly in four dimensions.

One of many examples where this model brought about conceptual advances for generic
quantum field theories is in understanding the structure of Feynman integrals \cite{ArkaniHamed:2010gh},
which is closely related to their analytic evaluation \cite{Henn:2013pwa}.
The three-loop planar Feynman integrals needed for the amplitude mentioned above were computed in ref. \cite{Caron-Huot:2014lda},
using a version of the differential equations method with improvements for integrals finite in four dimensions.
These formulas provide a fully analytic result for a three-loop four-point amplitude depending on three scales.

One exciting feature of the amplitude is that many of its physically interesting
limits are either exactly known, or governed by integrability. 
This is the case for the Regge limit, which at leading power is controlled by the anomalous
dimension of a cusped Wilson loop, a problem that is known to be integrable \cite{Correa:2012hh,Drukker:2012de}.
The amplitude is exactly known in the high-energy limit, and equal to the Bern-Dixon-Smirnov (BDS) ansatz \cite{Bern:2005iz,Drummond:2007au,Alday:2009zm}.
Moreover, the low-energy limit is described by an effective action,
and in the forward limit the amplitude can be interpreted as a total cross-section of producing
massive W bosons and other particles; an exact formula for that cross-section
at high energies will be conjectured below. Finally, it is possible to study threshold effects that are related
to bound states of a hydrogen-like system \cite{Caron-Huot:2014gia}.

In this paper, we study in detail the various physical limits mentioned above,
focusing in particular on the Regge limit.
It is well-known that in the planar limit it is described by a single power law, involving the gluon Regge trajectory,
which is a nontrivial function of the internal masses and momentum transfer.
Surprisingly, we will observe from the explicit three-loop 
amplitude that the first subleading power is also controlled by a single power law.
This is rather remarkable in light of the current understanding of subleading powers in the Regge limit,
where, for example, quark loops typically produce double logarithmic corrections
(see \cite{Bartels:2003dj,Kovchegov:2015pbl}),
In contrast, in this theory we find only a single power of logarithm per loop order.

In order to better understand this phenomenon and hopefully initiate a systematic expansion in the Regge limit,
we will make use of a special property of this model, wherein the Regge limit can be mapped to a limit of a massless internal leg.
This was used before to give an alternative definition of the Regge trajectory as the anomalous dimension of a cusped Wilson loop with a finite angle.
This kind of soft limit is at the moment under better theoretical control since it is conceptually closer to the limit of soft external momenta
studied in \cite{Laenen:2010uz,Luna:2016idw,Moult:2017rpl}.  We will make a proposal for an independent definition of the first subleading exponent, namely
as the anomalous dimension of a Wilson loop with a scalar inserted at the cusp.
We test this proposal by explicitly computing this anomalous dimension up to two loops.

The paper is organized as follows.
In section \ref{sec_introduction_amplitudes_limits}, we define the model and amplitudes under consideration. 
We discuss in detail the various physical limits and point out the all-loop structure expected in
some of them, using the one-loop result as a pedagogical example.
Section \ref{sec_subleading_regge} summarizes our observations about the Regge limit at next-to-leading power,
up to three loops. 
Then, in section \ref{sec_wilson_loop} we compute the anomalous dimension of a cusped Wilson loop
with a scalar insertion, and test our proposal that its anomalous dimension is equal to the second Regge trajectory
appearing at subleading power.
We present our conclusions in section \ref{sec_conclusions}.
The paper contains several Appendices with technical details.
Appendix \ref{sec_cross_section} collects our results for a total cross-section, and evidence for its conjectured
high-energy limit.  Appendix \ref{app_partial_waves} explains the use of dual conformal symmetry to conveniently parametrize the Regge expansion.
Appendix \ref{app_expansions} contains a detailed account of the analytic continuation and differential equation technology
needed to derive the various expansions of the three-loop scattering amplitude. 
In Appendix \ref{app_soft_current} we discuss the calculation of the Feynman integrals for the two-loop Wilson line calculation.

\section{Massive scattering amplitudes in $\mathcal{N}=4$ super Yang-Mills}  
\label{sec_introduction_amplitudes_limits}

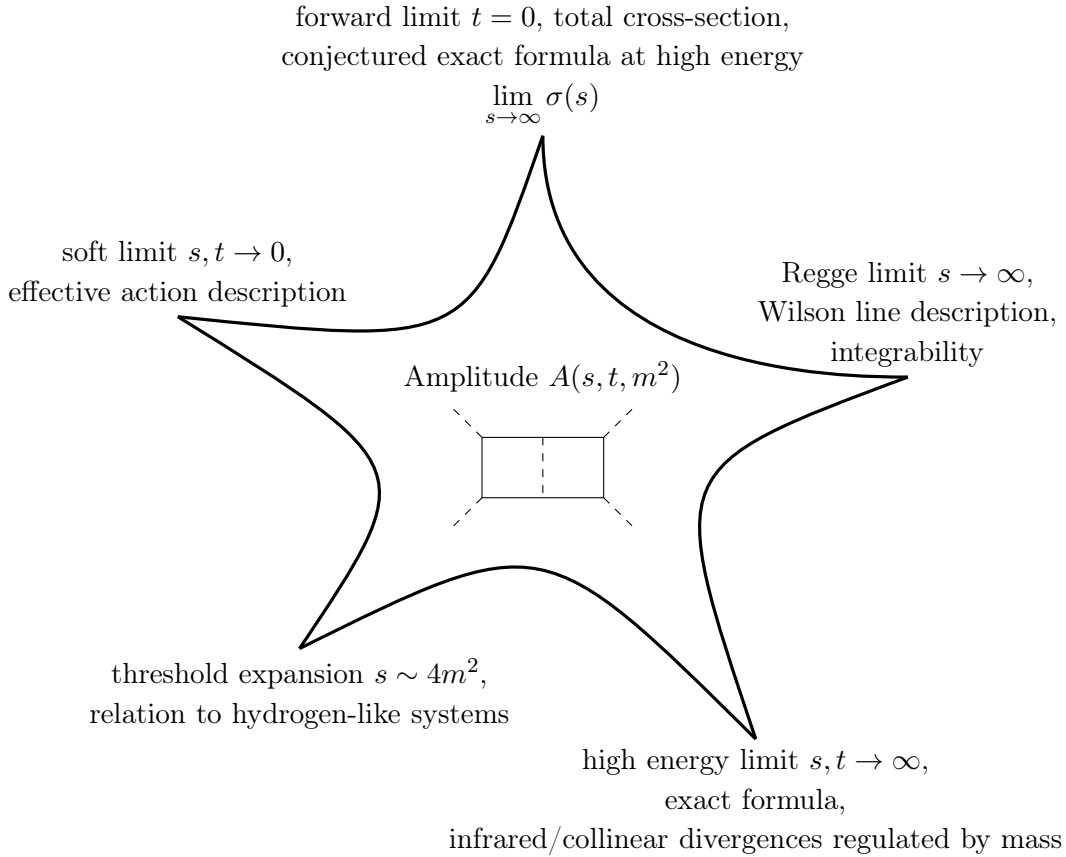
\begin{figure}[t]
\centering

     \begin{tikzpicture}[scale=0.8]

	  \coordinate (a) at (0,1.5) {};
	  \coordinate (b) at (6,-2.5) {};
	  \coordinate (c) at (3.5,-8.5) {};
	  \coordinate (d) at (-4,-7) {};
	  \coordinate (e) at (-6,-1.5) {};
	  \coordinate (corner) at (-1,-3.5) {};
	  
	  \draw[very thick] (a) .. controls (0,-2.5) and (5,-2.5) .. (b);
	  \draw[very thick] (b) .. controls (2,-4) .. (c);
	  \draw[very thick] (c) .. controls (0,-5) .. (d);
	  \draw[very thick] (d) .. controls (-2,-4) .. (e);
	  \draw[very thick] (e) .. controls (-1.2,-2) .. (a);
	  
	  \draw (corner) -- ++(2,0) -- ++(0,-1) -- ++(-2,0) -- ++(0,1);
	  \draw[dashed] (corner) -- ++(-0.5,0.5);
	  \draw[dashed] ($(corner) + (2,-1)$) -- ++(0.5,-0.5);
	  \draw[dashed] ($(corner) + (2,0)$) -- ++(0.5,0.5);
	  \draw[dashed] ($(corner) + (0,-1)$) -- ++(-0.5,-0.5);
	  \draw[dashed] ($(corner) + (1,0)$) -- ++(0,-1);
	  
	  \node at (a) [anchor = south, align=center] {forward limit $t=0$, total cross-section, \\ conjectured exact formula at high energy \\ ${\displaystyle \lim_{s \rightarrow \infty} \sigma(s)}$};
	  
	  \node at (b) [anchor = south, align=center] {Regge limit $s \rightarrow \infty$, \\ Wilson line description, \\ integrability};
	  
	  \node at (c) [anchor = north, align=center] {high energy limit $s,t \rightarrow \infty$, \\ exact formula, \\ infrared/collinear divergences regulated by mass};
	  
	  \node at (d) [anchor = north, align=center] {threshold expansion  $s \sim 4 m^2$, \\ relation to hydrogen-like systems};
	  
	  \node at (e) [anchor = south, align=center] {soft limit $s,t \rightarrow 0$, \\ effective action description};
	  
	  \node at (0,-2.5) [align=center] {Amplitude $A(s,t,m^2)$};
      
      \end{tikzpicture}
\caption{The scattering amplitude $A(s,t,m^2)$ has various physically interesting and overlapping
limits. In many of the latter, exact results are known or conjectured (e.g. high-energy limit), while other limits are known to be governed by integrability.}
\label{fig:amplitudes_introduction}
\end{figure}

\subsection{Setup and four-particle amplitudes}

We consider the ${\mathcal N}=4$ super Yang-Mills theory in the planar limit. We spontaneously break the $SU(N_c)$ gauge group to $SU(N_{c}-4) \times SU(4) \times U(1)$. 
In this way, in addition to the ``gluons'' of the unbroken $SU(N_{c}-4)$ part of the gauge group, we have massless bosons from the unbroken $SU(4)$ subgroup, a $U(1)$ photon, and massive $W$ bosons from the off-diagonal part. In what follows we will take $N_c$ large and discuss the leading term of the amplitudes.

As discussed in ref.~\cite{Alday:2009zm}, 
this allows us to consider color-ordered amplitudes $Y \bar{Y} \to Y \bar{Y}$.\footnote{One can also consider scattering of the $U(1)$ photons, as was done in ref. \cite{Bianchi:2015cta}.}
For further papers discussing various aspects of massive amplitudes on the Coulomb branch of $\mathcal{N}=4$ super Yang-Mills, see \cite{Schabinger:2008ah,Gorsky:2009dr,Boels:2010mj,CaronHuot:2010rj,Dennen:2010dh,Henn:2010bk,Henn:2010ir,Craig:2011ws,Kiermaier:2011cr,Henn:2011by,Correa:2012nk,Plefka:2014fta}.
Here the particle $Y$ is one of the off-diagonal generators of the unbroken SU(4) subgroup, lying above the diagonal; $\bar Y$ is then the Hermitian conjugate.
We choose $Y$ to be a complex scalar within the $\mathcal{N}=4$ supermultiplet, but since the setup preserves supersymmetry any other helicity choice would give an equivalent result.
An important motivation for considering such amplitudes is that they are natural from the AdS/CFT viewpoint \cite{Alday:2007hr}, and that they have a dual conformal symmetry \cite{Alday:2009zm}.

At tree-level, the result for the scattering amplitude is the same as in the unbroken theory,
\begin{align}\label{ampYYbartree}
  A_{Y \bar{Y}\rightarrow Y \bar{Y} }^{\rm tree} &= {-2\gYM^2} \frac{s}{t}
\end{align}
Amplitudes with other external states, such as gluons, are related to this one by supersymmetry. 
At loop level, at leading order in $N_c$, the interactions are mediated by massive W bosons, 
whose mass provides a natural infrared regularization. 
We define
\begin{align}\label{ampYYbar}
 A_{Y \bar{Y}\rightarrow Y \bar{Y}} &= A_{Y \bar{Y} \rightarrow Y \bar{Y}}^{\rm tree} \;
 M\left(\frac{4m^2}{-s},\frac{4m^2}{-t}\right)\,.
\end{align}
The amplitude $M$ can be expanded perturbatively in the coupling $g^2\equiv \gYM^2 N_c/(16 \pi^2)$, as 
\begin{align}\label{Mpert}
M= 1+ g^2 M^{(1)} + g^4 M^{(2)} +g^6 M^{(3)}+\mathcal{O}\left(g^8\right) \,.
\end{align}
The expression for the loop integrand of $M$ up to four loops 
was derived using unitarity cuts \cite{Bern:2010qa}. The loop integrals up to three loops were evaluated analytically in ref. \cite{Caron-Huot:2014lda}.
The main focus of this paper is to investigate the various limits discussed above, and to understand surprising structures appearing in them. We use the technology of ref. \cite{Caron-Huot:2014lda} to derive the expansions.  

In this section, we use the one-loop expressions as a pedagogical example, and point out the all-loop structure, whenever the latter is known.
The one-loop term $M^{(1)}$ of eq. (\ref{Mpert}) is given by a massive one-loop box integral,
which evaluates to (the form below is due to \cite{Davydychev:1993ut})
\begin{align}\label{I1exact}
M^{(1)} =&  -\frac{2 }{\buv} \Big\{ 
2 \log^2\left( \frac{\buv + \bu}{\buv + \bv} \right) + 
\log\left( \frac{\buv - \bu}{\buv + \bu} \right)  \log \left( \frac{\buv - \bv}{\buv + \bv} \right)
- \frac{\pi^2}{2} \nonumber\\
& + \sum_{i=1,2} \Big[ 
2 \, \Li_{2} \left( \frac{\beta_{i} - 1}{\buv + \beta_{i}} \right)
-2 \, \Li_{2} \left(- \frac{\buv - \beta_{i}}{\beta_{i} + 1} \right)
- \, \log^2 \left(\frac{\beta_{i} +1}{\buv + \beta_{i}} \right)
\Big]
 \Big\} \,. 
\end{align}
Here we introduced dimensionless variables
\begin{align}
u = \frac{4m^2}{-s}\,,\qquad \v=\frac{4m^2}{-t}\,,
\end{align}
and the following abbreviations,
\begin{align}
\bu=\sqrt{1+u}\,,\qquad \bv=\sqrt{1+v}\,,\qquad \buv=\sqrt{1+u+v}\,.  \label{def_beta}
\end{align}
The functions appearing in eq. (\ref{I1exact}) are examples of polylogarithms, with the dilogarithm defined as ${\rm Li}_{2}(x) = -\int_0^x \log(1-y)/y \,dy$. 
The above formulas are valid in the Euclidean region $u,v>0$. In order to continue to other regions, a small imaginary part has to be added to $s$ and $t$, according to the Feynman prescription.

As already mentioned in the introduction, the amplitude has several physically interesting limits, that we discuss presently,
as summarized in Fig.~\ref{fig:amplitudes_introduction}.

\subsection{Soft limit}
When $|s|,|t|\ll m^2$ (keeping $s/t$ fixed), the massive $W$ bosons can be integrated out, 
leading to a local effective action.
At tree-level, the massive $W$ bosons do not appear when scattering the light 
SU(4) particles, so that the scattering amplitude is the same as in the unbroken theory.
On the other hand, at loop level (and in the large $N_{c}$ limit), the light particles do not 
interact directly among themselves, but through a loop of massive $W$ bosons.
We have
\begin{align}
\frac{1}{s t} {M\left( \frac{4m^2}{-s},\frac{4m^2}{-t} \right)} = \frac{1}{st}-\frac{g^2}{6 m^4} + \mathcal{O}(1/m^6).
\end{align}
In this formula, the ${g^2}/{(6 m^4)}$ term is one-loop exact,
as predicted from non-renormalization theorems
(see ref.~\cite{Buchbinder:2001ui} and references therein). 

\subsection{Forward limit and total cross-section}

In the forward limit $t=0$, the optical theorem relates the imaginary part of the scattering amplitude of massless scalars
$Y \bar{Y} \longrightarrow Y \bar{Y}$ to the total cross-section of $Y, \bar{Y}$ producing a pair of massive W bosons, plus other particles. 
We have \cite{PeskinSchroeder}
\begin{equation}\label{totalcrossPS}
 \sigma_{Y\bar{Y}\rightarrow X}=\frac{1}{2 E_{ \rm cm} p_{ \rm cm}} \lim_{t \rightarrow 0} \, \IM(A) = \frac{1}{s} \lim_{t \rightarrow 0} \, \IM(A_{Y \bar{Y}\rightarrow Y \bar{Y}}) \; ,
\end{equation}
where $E_{\rm cm}=\sqrt{s}$ is the center of mass energy and $p_{\rm cm}=\sqrt{s}/2$ is the center of mass momentum of one particle. 
We have
\begin{equation}
 \lim_{t \rightarrow 0}  A_{Y \bar{Y}\rightarrow Y \bar{Y}}={-2 g_{YM}^2}\, \lim_{t \rightarrow 0}   \frac{s}{t} M \left(\frac{4m^2}{-s},\frac{4m^2}{-t} \right)
\end{equation}
In the Euclidean region $-s>0$ we find  
\begin{align}\label{forward1loop}
 \lim_{t \rightarrow 0} \frac{-m^2}{t} M^{(1)} =&\beta_u  \, {\log \left( \frac{\beta_u-1}{\beta_u+1} \right)}+2 
\end{align}
Analytically continuing to $s>4 m^2$ (taking into account the Feynman $i0$ prescription), and taking the imaginary 
part of eq. (\ref{forward1loop}), and using formula (\ref{totalcrossPS}), we arrive at
\begin{align}
 \sigma_{Y\bar{Y}\to X}=  \frac{2\pi g^2_{\rm YM}}{m^2} \beta_u+ \mathcal{O}(g^4) = \frac{32 \pi^3 g^2}{N_{c} m^2} \beta_u + \mathcal{O}(g^4) \,.
\end{align}
In Appendix \ref{sec_cross_section}, we compute this cross-section to the three-loop order,
and observe a simple pattern, which we argue allows us to propose an exact formula of its high energy limit:
\begin{align}
 \lim_{s\to \infty} \sigma_{Y \bar{Y}\rightarrow X} = 
 \frac{2\pi g^2_{\rm YM}}{m^2} B(g^2).
 \label{sigmatot}
\end{align}
Here $B(g^2)$ is the Bremsstrahlung function, given below in eq.~(\ref{Bremsstrahlungexact}) as an \emph{exact} function of the coupling.
It is striking, in particular, that the total cross-section for massless scalars or photons remains finite as $s\to \infty$.  
This is likely a consequence of working at the leading order in the large $N_{c}$ limit (which neglects, in particular,
the interactions within the unbroken SU(4) subgroup).
 
It is instructive to recall Pomeranchuk's theorem \cite{bookBogoliubov}, which states that
if the cross-section grows with energy, the cross-section for a
particle and its antiparticle will be asymptotically equal. The
hypothesis of the theorem is not satisfied: here the cross-section
does not grow with energy. Interestingly, the conclusion is also
maximally violated: the amplitude for the antiparticle process vanishes at this order: $\sigma_{YY\to X}=O(1/N_c^2)$.

\subsection{Threshold expansion}
Let us consider the amplitude close to the threshold $s=4m^2$ for producing a pair of $W$ bosons.
We expect the perturbative series to break down in the regime when the velocity $\bu\sim g^2$, for the following reason.
The produced $W$ bosons interact by exchanging massless gauge fields and scalars (and fermions), from the unbroken part of the gauge group,
which lead to an attractive $1/r$ potential. This causes the $W$ bosons to form non-relativistic Hydrogen-like bound states, which are exactly stable in the large $N_c$ limit.
Their binding energies are of order $\Delta E\sim mg^4$ at weak coupling.  While one cannot see these bound states in our fixed-order calculation,
one should still expected to see the perturbative series diverge when the kinetic energy becomes of this order. Recalling that $\bu=\sqrt{1-4m^2/s}$, this indeed translates to $\bu\sim g^2$.

Physically, the leading terms should originate from a nonrelativistic hydrogen-like system with the Hamiltonian in the center-of-mass frame\footnote{The potential is twice that coming from gauge boson exchange, due to the scalar exchange.}
\be
 H= \frac{p^2}{m} -\frac{\lambda}{4\pi r}\,.
\ee
The contribution of this system can in fact be computed analytically as a function of $g^2/\bu$ (see ref.~\cite{Beneke:2013jia}, eq.~(4.55)):
\be 
 \lim_{\bu\to 0^+}\frac{m^2{\rm Im} M(u,\frac{-4m^2}{t})}{-\pi g^2t} = \frac{4\pi^2g^2}{1-e^{-\frac{4\pi^2g^2}{\bu}}}
+ \mathcal{O}(\bu^2,g^2\bu,\ldots). \label{coulomb_resummation}
\ee
To all orders in $g^2$ this predicts the most singular term as the velocity $\bu\to 0$.
This resummation accounts for certain ladder graphs; these are the same graphs which govern the Regge limit.
There is in fact a very close connection between these two limits, as discussed in ref. \cite{Caron-Huot:2014gia}.
Higher order corrections to eq. (\ref{coulomb_resummation}) should be interpreted as relativistic and many-body corrections to the
Coulomb Hamiltonian. 

\subsection{High energy limit}
We can take $s,t$ to be much larger than the mass, $m^2/s \to 0, m^2/t \to 0$, with $s/t$ fixed.
In this case, the mass serves as an infrared and collinear regulator, leading to double logarithms of the small mass.
Expanding eq. (\ref{I1exact}) in this limit, we obtain
\begin{align}\label{M1highenergy}
{M\left( \frac{4m^2}{-s},\frac{4m^2}{-t} \right)}  = 1 + g^2  \left[ -2 \log\left( \frac{m^2}{-s} \right)  \log\left( \frac{m^2}{-t} \right)  + \pi^2   \right] + \mathcal{O}(g^4)\,.
\end{align}

It was argued \cite{Alday:2009zm}, based on anomalous dual conformal Ward identities originally 
derived for Wilson loops \cite{Drummond:2007au}, 
that the four-point amplitude should have the following exact form,
\begin{align}\label{Mhighernergy2}
\log M\left( \frac{4m^2}{-s},\frac{4m^2}{-t} \right) =& \frac{\gamma(g)}{8} \left[ -2 \log\left( \frac{m^2}{-s} \right)  \log\left( \frac{m^2}{-t} \right) + \pi^2 \right]  \nonumber \\& + \tilde{\mathcal{G}}_{0}(g)  \left[  \log\left( \frac{m^2}{-s} \right)  +\log\left( \frac{m^2}{-t} \right) \right] + \tilde{c}(g)   + \mathcal{O}(m^2) \,,
\end{align}
where $\gamma(a)$ is the light-like cusp anomalous dimension \cite{Korchemskaya:1992je,Beisert:2006ez},
$\tilde{\mathcal{G}}_{0}$ is a collinear anomalous dimension, and $\tilde{c}$ a coupling-dependent constant.
Eq. (\ref{Mhighernergy2}) can be viewed as a mass-regulated version of the BDS ansatz \cite{Bern:2005iz}, which was originally formulated within dimensional regularization. 

The small mass limit and eq. (\ref{Mhighernergy2})  were studied previously using Mellin-Barnes techniques
in refs. \cite{Henn:2010bk,Henn:2010ir}. In the preceding sections, we derived analytic formulas for $M$ up to three loops.
As a check, we reproduced  eq. (\ref{Mhighernergy2}) to that order by taking the small mass limit of our formulas.
For reference the coefficients are $\gamma(g^2)=8g^2-16\zeta_2g^4+176\zeta_4g^6$; 
$\tilde{\mathcal{G}}_{0}=-4\zeta_3g^4+g^6(36\zeta_5-8\zeta_2\zeta_3)$;
$\tilde{c}(g^2)=3g^4\zeta_4-g^6(50\zeta_6+16\zeta_3^2)$.

The fact that the logarithm of the amplitude does not grow faster than $\log(s)$ is consistent with behaviour expected in the Regge limit $s\gg m^2$, that will be discussed presently.
The fact that eq. (\ref{Mhighernergy2}) contains only a single logarithm (and no further $s$ dependence) means that $M$ is Regge exact (in the small mass limit).\footnote{The property of Regge-exactness was observed in the dimensionally-regularized massless case in refs. \cite{Drummond:2007aua,Naculich:2007ub}.}

\subsection{Regge limit}

The leading term in the Regge limit $s \gg m^2,t$, up to power corrections, has been discussed in
refs.~\cite{Henn:2010bk,Henn:2010ir}.
It is given by a single power law,
\begin{align}
 \lim_{s\to\infty} M\left(\frac{4m^2}{-s},\frac{4m^2}{-t}\right) =& \tilde{r}_0(t) \left( \frac{-s}{m^2}\right)^{j_0(t)+1} + \mathcal{O}(1/s)\,. \label{LO_Regge_behavior}
\end{align}
The leading terms are given by
 \begin{align} \label{LO_trajectory1}
  & j_0 = -1 +2 g^2 \phi \tan \frac{\phi}{2} + \mathcal{O}\left(g^4\right) \, , & & \tilde{r}_0=1+ \mathcal{O}\left(g^2\right),  \end{align}
 where
  \begin{align} 
 t = 4 m^2 \sin^2 \frac{\phi}{2}\,.
 \end{align} 
We note that in planar $\mathcal{N}=4$ super Yang-Mills, the Regge trajectory is equal to the angle-dependent cusp anomalous dimension \cite{Henn:2010bk},
\begin{align} \label{Relation_GammaCusp-j0}
j_0(t) + 1= - \Gamma_{\rm cusp}(\phi) \,.
\end{align}
The angle-dependent cusp anomalous dimension is an extremely interesting quantity in its own right. In QCD it is known up to three loops \cite{Grozin:2015kna}, while in $\mathcal{N}=4$ sYM theory it was computed up to three loops in \cite{Correa:2012nk} and in the planar limit up to four loops in \cite{Henn:2013wfa}. Furthermore it is controlled by integrability \cite{Correa:2012hh,Drukker:2012de} in $\mathcal{N}=4$ sYM theory. We refer the interested reader to \cite{Correa:2012nk,Grozin:2015kna} for a discussion of its various properties.
Here we wish to point out that its small angle limit is known exactly \cite{Correa:2012at},
\begin{align}\label{Bremsstrahlungexact}
 \Gamma_{\rm cusp}(\phi) = - \phi^2 \, B(g^2)\,, \qquad B(g^2)=
 \frac{1}{4\pi^2} \frac{\sqrt{\lambda}I_2(\sqrt{\lambda})}{I_1(\sqrt{\lambda})}  \approx g^2 -\frac23 \pi^2 g^4+ \frac23 \pi^4 g^6 + \ldots \,.
\end{align}
Here $\lambda = g_{\rm YM}^2 N_{c}$, and $I_{1}$ and $I_{2}$ are Bessel functions.

Computing the three-loop Regge limit (\ref{LO_Regge_behavior}) of the amplitude, as described in Appendix \ref{app_expansions},
we computed $j_{0}(t)$, and hence $\Gamma_{\rm cusp}(\phi)$, to the three-loop order.
In this way, we reproduced the result of ref.~\cite{Correa:2012nk}. For the explicit expressions we refer to that reference.

\subsection{Discussion}

In summary, we studied a scattering amplitude of four complex scalars in $\mathcal{N}=4$ sYM theory with a spontaneously broken gauge group,
which is known as an analytic function of the variables $u=4m^2/(-s)$ and $v=4m^2/(-t)$ up to three loops \cite{Caron-Huot:2014lda}.
We studied several kinematic limits. To obtain these limits we used techniques for differential equation and calculated the amplitude in an (asymptotic) expansion, in principle to any order in the expansion parameter. 
For convenience of the reader, we collet our results in a computer-readable ancillary file.
Interestingly, we find that the leading terms of most of the above limits are in principle known to all loop orders, or are controlled by an integrable model.

In the case of massless amplitudes, a systematic expansion around the collinear limit could be found and described via integrability \cite{Alday:2010ku,Basso:2013vsa}.
The above observations nurture the hope that something similar can be done for massive amplitudes, at least in one of the above limits. 
Focusing on the first subleading term in the Regge limit, we will find a remarkably simple structure. In the remainder of this paper, we discuss these findings in detail.


\section{Regge expansion using dual conformal partial waves}
\label{sec_subleading_regge}

The Regge limit is a likely candidate around which one can hope to build a systematic expansion.
When discussing power-suppressed terms, however, the choice of variable used to express the leading term (\ref{LO_Regge_behavior})
becomes important.  In this section we will use symmetries to derive a good parametrization of the limit,
and we will find that the first power-correction is then controlled by a single independent power.

\subsection{Dual conformal partial wave analysis}

A simple improvement over expanding in inverse powers of $1/s$ in the Regge limit at fixed $t$
is to use instead the partial wave expansion, where each power gets upgraded to a Regge pole.
Each Regge pole contributes proportional to a Legendre function, producing
an asymptotic expansion of the amplitude in the form (see eq.~(A.12) of \cite{Donnachie:2002en} or (2.9.6) of \cite{Collins:1977jy}):
\be
 A(s,t) \simeq \sum_{j_n(t)} \tilde{c}_n(t) Q_{-j_n(t)-1}(\cos\theta_3), \quad\mbox{with}\quad \cos\theta_3 =  1+\frac{2s}{t}\,, \label{SO(3)}
\ee
where $j_n(t)$ denote the Regge trajectories and $Q_{-j-1}$ are associated Legendre functions.
In the Regge limit these tend to\footnote{The right-hand side shows the expansion of:
$\frac{{-}\tan(\pi j)\Gamma(j+1)}{\sqrt{\pi}2^j\Gamma(j+\frac12)}Q_{-j-1}(z) =z^j {}_2F_1(-\tfrac{j}2,\tfrac{1-j}{2},\tfrac12-j,1/z^2)$.}
\be \lim_{s\to\infty} Q_{-j-1}(\cos\theta_3) \propto 
 \left(\frac{2s}{t}\right)^j+j\left(\frac{2s}{t}\right)^{j-1} + \frac{j(j-1)^2}{2j-1} \left(\frac{2s}{t}\right)^{j-2}+ \ldots \label{Q regge}
\ee
This highlights how each Regge pole resums an infinite tower of inverse powers of $s$.
The Legendre functions which control each tower originate physically from
the SO(2,1) Lorentz subgroup which preserves the spacelike momentum exchanged in the $t$-channel.
For this reason, the expansion (\ref{SO(3)}) generally contains fewer independent coefficients than the straightforward $1/s$ expansion.
This is analogous to how conformal blocks are used to resum descendant operators in conformal field theories.

The $\mathcal{N}=4$ sYM amplitude that we consider benefits from a larger SO(4) or SO(3,1) dual conformal symmetry.
It was identified in \cite{Caron-Huot:2014gia} as a relativistic version of the Laplace-Runge-Lenz symmetry of the
hydrogen-like bound states which appear in intermediate states of the amplitude.
This symmetry implies further relations among the Regge trajectories. 

Using the embedding formalism to work out the action of the dual conformal symmetry on the kinematical invariants,
and the detailed form of SO(4) Legendre functions,
a symmetry-improved expansion around the Regge limit is derived in Appendix \ref{app_partial_waves}:
\be
\lim_{Y\to 0} \frac{1+Y}{1-Y} M = \sum_{n=0}^\infty r_n(t) Y^{-j_n(t)-1}\,,\label{SO4_final}
\ee
where the variable, which vanishes in the Regge limit $s\to\infty$, is (recall eq.~(\ref{def_beta}) for our notations)
\be
 Y = \frac{\buv-\bv}{\buv+\bv} \equiv e^{i\theta} \quad\mbox{where} \quad \cos\theta=1+\frac{2s}{t} - \frac{s}{2m^2}. \label{Y angle}
\ee
Again each term behaves in the Regge limit like a power of $s$: $Y^{-j-1}\propto s^{j+1}$,
however an intricate tower of subleading powers, similar to but distinct from eq.~(\ref{Q regge}), is associated with each SO(4) Regge trajectory $j_n(t)$.

The angle (\ref{Y angle}) is distinct from the usual scattering angle between the
two external massless photons (see eq.~(\ref{SO(3)})), due to the $-s/(2m^2)$ term.
This makes the angle real in the physical region of $t$-channel scattering when $0<t<4m^2$ and $-t<s<0$.
It is hyperbolic above the massive continuum $t>4m^2$, or whenever $|s|$ is sufficiently large,
as is assumed in the Regge limit formula (\ref{SO4_final}).

As an application of this formula, we have subtracted from the three-loop amplitude described in the preceding section
its leading behavior: $r_0(t) Y^{-j_0(t)-1}$, which is known to exponentiate to a single power law (with exponent previously given
in ref.~\cite{Correa:2012nk, Gromov:2015dfa}).
Amazingly, looking at the remainder, we find that the first subleading power also exponentiates!  That is, we find to three loops that
the logarithm of the remainder is linear in $\log Y$.  We can thus write
\be
 \lim_{Y\to 0} \frac{1+Y}{1-Y} M = r_0(t) Y^{-j_0(t)-1}+ r_1(t) Y^{-j_1(t)-1} + O(Y^2)
\label{Regge NLO}
\ee
where $j_0{\approx}-1$ and $j_1{\approx}-2$.
This is rather remarkable: in principle the first subleading term could have been a sum of multiple power laws.
In fact, had we used the straightforward $1/s$ expansion, or the usual SO(3) Regge pole expansion,
we would have (artificially) found at subleading order two distinct power laws,
with one exponent mysteriously equal to the leading exponent minus one, thus hinting at the presence of a hidden symmetry.

Since we have only one single subleading power, taking the limit of the three-loop amplitude of ref.~\cite{Caron-Huot:2014lda}
using the method detailed in Appendix \ref{app_expansions},
allows us to extract its exponent to three loops.
Defining 
\be
 x=\frac{\bv-1}{\bv+1}, \qquad \xi =\frac{1}{\bv} = \frac{1-x}{1+x}, 
\ee
our result for the Regge trajectory is:
\begin{align}
j_1=& -2-4g^2 + g^4 \left[-\frac{1}{\xi } \left(4 \zeta _2 H_1+\frac{H_1^3}{6}\right)-8 \xi H_1 +2 H_1^2 +16 \left(\zeta _2+1\right)\right] \label{j1_threeloops} \\
& g^6 \Bigg[  32 H_{1,2}-16 \left(4 \zeta _2-6 \zeta _3+15 \zeta _4+8\right)-12 \left(2 \zeta _2+1\right) H_1^2-\frac{5 H_1^4}{6}+\frac{8 H_1^3}{3} \nonumber \\
&\quad+ \xi \left( 48 \left(\zeta _2+2\right) H_1+\frac{14 H_1^3}{3}-16 H_1^2-64 H_2 \right) -8 \xi^2 H_1^2 \nonumber \\
&\quad+ \frac{1}{\xi} \left( -8 H_{1,1,2}+\frac{2}{3} \left(4 \zeta _2+1\right) H_1^3+8 \left(2 \zeta _2-3 \zeta _3+11 \zeta _4\right) H_1+\frac{H_1^5}{20}-\frac{H_1^4}{6} \right) \Bigg]\, , \nonumber
\end{align}
where $H$ are harmonic polylogarithms \cite{Remiddi:1999ew,Maitre:2005uu} with argument $1-x^2$.
The residue starts as $r_1 = 2 + 8 g^2\left(2 H_{-1}(x)+2 H_1(x)-1\right)+\mathcal{O}(g^4)$.

We also looked at the next powers in the expansion.  We find that there is never any double logarithm, that is, never more than one power of logarithm per loop order.
To leading logarithm order, the sub-sub-leading term in the amplitude is given as
\begin{align}
\label{subsubleading}
\frac{1+Y}{1-Y} \, M(u,Y) \bigg|_{Y^2} =& \,
Y^2 \Bigg[ 2+8 g^2 \log(Y)\frac{\xi^2-1}{\xi^2} -8 g^4 \log^2(Y)  \frac{\left(2 \xi ^4-6 \xi ^2+5\right)}{\xi ^2 \left(1-\xi ^2\right)} \nonumber \\
					& \qquad -64 g^6 \log ^3(Y)\frac{ \xi^4-4 \xi^2  + 5}{3 \xi ^2 \left(1-\xi^2\right)}\Bigg].
\end{align}
One can show that this not the exponential of a single power.

For convenience, the full three loop expansions of $j_0$, $j_1$, $r_0$, $r_1$ and sub-subleading amplitudes
are recorded in an ancillary file attached to the arXiv submission of this paper.

\subsection{From the Regge limit to Wilson lines with a cusp}
\label{sec:fromReggetoWilson}

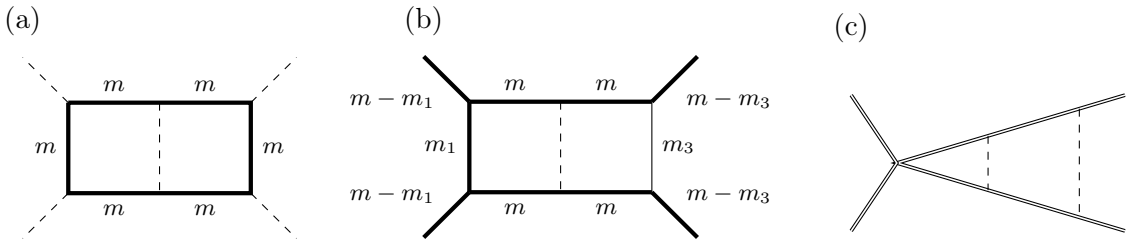
\begin{figure}[t]

 \begin{minipage}{.27\textwidth} 
     \centering
    \begin{tikzpicture}[scale=1.2]
	  
	  \node at (-0.5,0.9) {(a)};

	  \coordinate (a11) at (0,0) {};
	  \coordinate (a12) at (1,0) {};
	  \coordinate (a13) at (2,0) {};
	  \coordinate (a21) at (0,-1) {};
	  \coordinate (a22) at (1,-1) {};
	  \coordinate (a23) at (2,-1) {};
	 
	  
	  \draw[ultra thick] (a11) -- node[anchor=south, midway] {\footnotesize $m$} (a12) -- node[anchor=south, midway] {\footnotesize $m$} (a13) -- node[anchor=west, midway] {\footnotesize $m$} (a23) -- node[anchor=north, midway] {\footnotesize $m$} (a22) -- node[anchor=north, midway] {\footnotesize $m$} (a21) -- node[anchor=east, midway] {\footnotesize $m$} (a11) {};
	  \draw[dashed] (a12) -- (a22);
	  \draw[dashed] (a11) --  ++(-0.5,0.5);
	  \draw[dashed] (a13) --  ++(+0.5,0.5);
	  \draw[dashed] (a21) --  ++(-0.5,-0.5);
	  \draw[dashed] (a23) --  ++(+0.5,-0.5);
	  
    \end{tikzpicture}
  \end{minipage}     
  \hfill  
 \begin{minipage}{.4\textwidth} 
     \centering
    \begin{tikzpicture}[scale=1.2]
    
    	  \node at (-0.5,0.9) {(b)};

	  \coordinate (a11) at (0,0) {};
	  \coordinate (a12) at (1,0) {};
	  \coordinate (a13) at (2,0) {};
	  \coordinate (a21) at (0,-1) {};
	  \coordinate (a22) at (1,-1) {};
	  \coordinate (a23) at (2,-1) {};
	 
	  
	  \draw[ultra thick] (a11) -- node[anchor=south, midway] {\footnotesize $m$} (a12) -- node[anchor=south, midway] {\footnotesize $m$} (a13);
	  \draw[ultra thick] (a23) -- node[anchor=north, midway] {\footnotesize $m$} (a22) -- node[anchor=north, midway] {\footnotesize $m$} (a21);
	  \draw[ultra thick] (a11) -- node[anchor=east, midway] {\footnotesize $m_1$} (a21);
	  \draw (a13) -- node[anchor=west, midway] {\footnotesize $m_3$} (a23);
	  \draw[dashed] (a12) -- (a22);
	  \draw[ultra thick] (a11) -- node[anchor=north east] {\footnotesize $m-m_1$} ++(-0.5,0.5);
	  \draw[ultra thick] (a13) -- node[anchor=north west] {\footnotesize $m-m_3$} ++(+0.5,0.5);
	  \draw[ultra thick] (a21) -- node[anchor=south east] {\footnotesize $m-m_1$} ++(-0.5,-0.5);
	  \draw[ultra thick] (a23) -- node[anchor=south west] {\footnotesize $m-m_3$} ++(+0.5,-0.5);
	  
    \end{tikzpicture}
     \end{minipage} 
     \hfill    
  \begin{minipage}{.27\textwidth} 
     \centering
        \begin{tikzpicture}[scale=1.2]
        
          \node at (-0.5,1.5) {(c)};

	  \coordinate (cusp) at (0,0) {};
	  \coordinate (a11) at (1,0.3) {};
	  \coordinate (a12) at (2,0.6) {};
	  \coordinate (a13) at (2.5,0.75) {};
	  \coordinate (a21) at (1,-0.3) {};
	  \coordinate (a22) at (2,-0.6) {};
	  \coordinate (a23) at (2.5,-0.75) {};
	  \coordinate (e11) at (-0.5,0.75) {};
 	  \coordinate (e21) at (-0.5,-0.75) {};

	 \draw[double] (a13) -- (a12) -- (a11) -- (cusp) -- (a21) -- (a22) -- (a23);
	 \draw[double] (e11) -- (cusp) -- (e21);
	  \draw[dashed] (a11) -- (a21);
	  \draw[dashed] (a12) -- (a22);
	  
	  
    \end{tikzpicture}
  \end{minipage} 
\caption{The amplitude on the left, with four massive $W$ bosons (thick lines)
running outside the loop, is equivalent, through eq.~(\ref{cross ratios}), to an amplitude
with unequal-mass bosons.  In a limit equivalent to the Regge limit, one of the masses
go to zero, revealing infrared divergences within the associated cusp.}
\label{fig:Amplitude_to_Wilson}
\end{figure}

While we currently lack a systematic effective field theory framework to characterize
the subleading powers (\ref{Regge NLO}) in the Regge limit, to make progress we will use a map
to an equivalent problem involving a cusped Wilson line, following \cite{Henn:2010bk,Correa:2012nk,Caron-Huot:2014gia}.
The idea is to generalize the Higgs symmetry breaking pattern by further breaking SU(4) down to U(1)${}^4$,
thus allowing a distinct mass $m_i$ for the W bosons.
The amplitude still depends only on six-dimensional dot products of the vectors (\ref{Yi}),
which give rise to two independent cross-ratios, now equal to \cite{Wick:1954eu,Cutkosky:1954ru}:
\be
 u = \frac{4m_1m_3}{-s+(m_1-m_3)^2}, \qquad v= \frac{4m_2m_4}{-t+(m_2-m_4)^2}.
 \label{cross ratios}
\ee
These reduce, in the equal-mass case, to our previous definitions, e.g~ $u=\frac{4m^2}{-s}$, see Fig.~\ref{fig:Amplitude_to_Wilson} (a).
That the amplitude $M$ depends only on these two cross-ratios has an implication which is familiar
in the non-relativistic limit: the cross-ratios then depend only on the kinetic energy divided by the reduced mass.
Most important for us, will be the fact that the Regge limit $s\to \infty$ of this amplitude, is equivalent to the massless limit $m_3\to 0$.
See Fig.~\ref{fig:Amplitude_to_Wilson} (b), where we also set $m_2=m_4 = m$ for simplicity.

\begin{figure}[t]
\centering

     \begin{tikzpicture}[scale=1.5]

	  \coordinate (cusp) at (0,0) {};
	  \coordinate (InW) at (-1,-1) {};
	  \coordinate (OutW) at (1,-1) {};
	  \coordinate (InPhi) at (0,-1) {};
	  \coordinate (le) at (-0.5,-0.5) {};
	  \coordinate (ri) at (+0.5,-0.5) {};
	  \coordinate (aux) at (+0.4,0.4) {};
	  
	  \draw[Wilson] (InW)  --  node[pos=0, anchor = north] {$v_1$} (le) -- node[pos=0, anchor = south east] {$C_1$} (cusp) -- node[pos=1, anchor = south west] {$C_2$} (ri) -- node[pos=1, anchor = north] {$v_2$} (OutW) {};	  
      
          \draw (cusp) -- (aux);
	  \draw (-45:0.35) arc [start angle=-45, end angle=45, radius=0.35];
	  \node (NN) at (0.2,0) {$\phi$};
      
      \end{tikzpicture}
\caption{Wilson line with two straight line segments forming a cusp.}
\label{fig:Wilson_Cusp}
\end{figure}
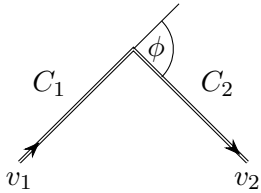

Logarithms in such a massless limit are naturally associated with soft quanta coupled to Wilson lines with a cusp geometry, see Fig.~\ref{fig:Amplitude_to_Wilson} (c).
This was used in \cite{Correa:2012nk} to obtain the 3-loop cusp anomalous dimension from the leading power in the Regge limit of the amplitude.
It was used in the other direction in \cite{Caron-Huot:2014gia}, to obtain the bound state spectrum of hydrogen-like states from the latter
(and further analyzed at strong coupling \cite{Espindola:2016afe}).  
Using the same map for the first subleading power, the subleading Regge trajectory $j_1$, see eq. (\ref{Regge NLO}) can thus be used to predict
that a Wilson line with a cusp has an excitation with scaling dimension:
\be
 \Gamma_{{\rm cusp},\Phi}(\phi) \equiv -j_1-1 = 1+4g^2+{\mathcal{O}}(g^4), \label{mainconjecture}
\ee
where $j_1$ is given to three loops by eq.~(\ref{j1_threeloops}).
The variable $x$ used in that expression can be expressed in terms of the cusp
angle $\phi$ between $p_3$ and $p_4$ shown in Fig.~\ref{fig:Wilson_Cusp} as
\be
 x=e^{i\phi}, \qquad i\xi = \tan\tfrac\phi2.
\ee

In summary, the fact that the first power correction in the Regge limit exponentiates
is remarkable and strongly suggests the existence of a systematic expansion with a nice structure. In the next section, we make a proposal for how to determine $j_1(t)$,
or equivalently the scaling dimension $\Gamma_{{\rm cusp},\Phi}(\phi)$, in terms of an effective field theory calculation.


\section{Renormalization of Wilson lines with operator insertions}
 \label{sec_wilson_loop}

In this section we show perturbatively up to two loops that the exponent $j_1$ in the power suppressed term in the Regge limit can be identified as the scaling dimension of a cusped Wilson loop with a scalar inserted at the cusp point. 

\subsection{Cusped Wilson line}

We recall that we found for the exponent $j_0$ of the leading term in the Regge limit the relation $j_0+1=-\Gamma_{\mathrm{cusp}}$, where $\Gamma_{\mathrm{cusp}}$ is the anomalous dimension of a Wilson loop operator $W_{\mathrm{cusp}}$ with a cusp; see eq. (\ref {Relation_GammaCusp-j0}). To define $W_{\mathrm{cusp}}$ properly, we consider the Maldacena-Wilson loop operator \cite{Maldacena:1998im}
\begin{equation} 
W[C]  = P \exp \left[ i g_{\rm YM} \int_{C} d{x}_{\mu} A^{\mu} + g_{\rm YM} \, \int_{C} |dx|  \Phi   \right] \,,
 \end{equation}
where $\Phi$ is one of the six scalars.
The fields are taken to be in the adjoint representation of the gauge group\footnote{Note that there is no trace in the definition, as the Wilson lines extend to infinity.}.
In the following, we will work in the large $N_c$ limit. The contour consists of two straight line segments $C_{1} = \{ \tau v_1^\mu| \tau \in [-\infty, 0] \}$ and $C_{2} =  \{ \tau v_2^\mu| \tau \in [ 0,\infty] \}$, which form a cusp at the origin. Note that the directions are chosen such that $v_1$ is incoming and $v_2$ is outgoing; see Fig.~\ref{fig:Wilson_Cusp}. The cusp angle is defined as
\begin{align}
\cos(\phi) = v_1 \cdot v_2 = \frac{1}{2} \left(x+\frac{1}{x} \right)\,, \qquad  v_1^2=v_2^2=1 \, .
\end{align}
The Wilson loop operator we are interested in is then simply given by 
\begin{equation}
W_{\rm cusp}=W[C_1] W[C_2] \, .  \label{Wcusp}
\end{equation}
It is invariant under the interchange $v_1 \leftrightarrow - v_2$, which follows from the definition of the contours.

The operator (\ref{Wcusp}) renormalizes multiplicatively: $W_{\rm cusp}^{\rm ren}  = Z^{-1}_{\rm cusp} W_{\rm cusp}$,
where the L.H.S. is finite \cite{Polyakov:1980ca, Gervais:1979fv, Dotsenko:1979wb, Arefeva:1980zd, Brandt:1981kf, Dorn:1986dt, Korchemsky:1987wg}. In a conformal field theory, such as $\mathcal{N}=4$ sYM, the renormalization
factor $Z_{\rm cusp}$ has the following form in $D=4{-}2\eps$ dimensions, see e.g. \cite{Dixon:2008gr},
\begin{align} \label{Z_cusp}
\log Z_{\rm cusp} = - \sum_{L \ge 1}  \frac{(g^2)^{L}}{2 L \epsilon}\,\Gamma_{\rm cusp}^{(L)} \, ,
\end{align}
where $\Gamma_{\rm cusp}(\phi) = \sum_{L\ge 1} (g^2)^{L} \Gamma_{\rm cusp}^{(L)} $ is the (angle-dependent) cusp anomalous dimension.

To motivate which subleading power operators to consider, let us discuss schematically what might be anticipated
from a systematic analysis of the massless limit $m_3\to 0$ of the preceding subsection,
in the framework of heavy quark effective theory (HQET).
One would start by integrating out the heavy (internal) $W$ bosons, which should reduce the $2\rightarrow 2$ amplitude
at leading power to a matrix element of local operators which creates a pair of heavy (scalar) quarks:
\be
q^\dagger(v_2)q(v_1). \label{HQET}
\ee
Taking the matrix element in a state with two heavy particles,
the propagator for the HQET fields $q$ simply produce Maldacena-Wilson lines, thus recovering $W_{\rm cusp}$.
Schematically, two kinds of power corrections are then expected: either from higher-dimension operators in the HQET Lagrangian,
or from higher-dimensional corrections to the local operator.
The former adds operator insertions along the Wilson lines, whereas the latter adds fields or derivatives strictly at the cusps.
For the present analysis, we shall assume that the leading power corrections come entirely from cusp insertions, and ignore the corrections to the Wilson lines.

\subsection{Operator mixing and renormalization}

For the first power correction, we are led to consider local HQET operators with mass dimension one higher
than (\ref{HQET}), for example $q^\dagger \Phi q$.
As we ignore corrections to the HQET Lagrangian, the heavy quark fields become again simply Maldacena-Wilson lines,
and we can treat this as a scalar insertion in $W_{\rm cusp}$.
Since the small mass $m_3$ is controlled by the Higgs mechanism, it should be better viewed as a property of the state rather than of the operator,
and in the free theory the scalar $\Phi$ simply becomes its expectation value $m_3$.
The considered operator can in principle mix with any other which has the same mass dimension and Lorentz indices.
The only gauge invariant operator built from $\mathcal{N}=4$ sYM fields fulfilling these criteria involve derivatives within the plane of the cusp
(and therefore total derivatives).
We are thus led to consider the two sYM operators:
\begin{align}
\partial{W}_{\rm cusp} =& \, 
\frac{i}{g_{\rm YM}}(v_2-v_1)_\mu \partial^\mu \, W_{\rm cusp}\,, \label{Def_W_cusp_mod}\\
W_{{\rm cusp}, \Phi} =& \, W[C_1] \, \Phi(x) \, W[C_2] \,.
\end{align} 
The relative sign in (\ref{Def_W_cusp_mod}) comes from the symmetry under $v_1 \leftrightarrow - v_2$.

In the case where the scalars which couple to the Wilson line are orthogonal to the scalars inserted at the cusp, the anomalous dimension is known from integrability \cite{Gromov:2015dfa}.
Here, however, these scalars are the same. This setup was considered by Alday and Maldacena in \cite{Alday:2007he}, where they computed the anomalous dimension at one loop for the straight line case. We extend the calculation to two loops with a dependence on the cusp angle. 

In general both operators can mix at loop level. In order to resolve this mixing, we will consider suitable correlation functions of these operators.
Since $\langle 0| \, W_{{\rm cusp},\Phi} | 0 \rangle$ has no tree-level contribution, we find it more convenient to consider the correlator with an additional
scalar $\Phi(p_3)$. We take the latter to be on-shell, $p_3^2=0$, so that the correlators are gauge independent. 

At tree-level, we find
\begin{align}
\langle 0| \, \partial W_{{\rm cusp}} \, |\Phi(p_3)  \rangle &= \frac{\big[(v_1-v_2){\cdot}p_3\big]^2}{(v_1{\cdot}p_3)(v_2{\cdot}p_3)} + {\mathcal{O}}(g^2)
 = \frac{(s_1+s_2)^2}{s_1 s_2}+{\mathcal{O}}(g^2) \,, \\
\langle 0| \, W_{{\rm cusp},\Phi} \, |\Phi(p_3) \rangle &=  1 + {\mathcal{O}}(g^2) \,.
\end{align}
These correlators depend on the cusp angle $\phi$, as well as on the two  invariants $s_1=-2 v_1 \cdot p_3$ and  $s_2=+2 v_2 \cdot p_3$.
In the following, the different momentum-dependence of the correlators, together with the fact that the (ultraviolet) renormalization matrix must be independent
of the external momentum $p_3$, will allow us to resolve the operator mixing.

Moreover, since $\partial {W}_{\rm cusp}$ is a derivative of the lower dimensional operator $W_{\textrm{cusp}}$, it renormalizes multiplicatively with the same renormalization factor, and we thus expect the mixing matrix to be triangular.
Taking also into account that operator mixing only appears at the loop level, we have the following structure of the renormalization matrix for $\vec{W} = \{ \partial W_{{\rm cusp}},  W_{{\rm cusp},\Phi} \} $,
\begin{equation} \label{Mix_Mat}
 \mathbf{Z}= \left( \begin{array}{cc}
			Z_{\rm cusp} & 0  \\
			Z_{\rm mix}  & Z_{{\rm cusp},\Phi} \end{array} \right) \,,
			\end{equation}
			with
			\begin{equation}
 Z_{\rm mix} = g^2 \frac{Z_{\rm mix}^{(1)}}{\epsilon}+\mathcal{O}(g^4) \, , \qquad Z_{{\rm cusp},\Phi} =1+ g^2 \frac{Z_{{\rm cusp},\Phi} ^{(1)}}{\epsilon}+\mathcal{O}(g^4) \, .
\end{equation}
In the following, we show up to two loops that $\mathbf{Z}$ is diagonal, i.e. that $W_{{\rm cusp},\Phi} $ renormalizes multiplicatively up to that loop order.  
This accidental vanishing of $Z_{\rm mix}$ might be related to the enhanced (dual conformal) symmetry of our setup, which we are not exploiting in the present calculation.

The correlators not only have UV divergences coming from the cusp and the operator insertion, but also soft and collinear divergences from the on-shell scalars. The latter can be renormalized with a common IR Z-factor
\begin{equation} \label{Log_ZIR}
 \log\left(Z_{\rm IR}^{-1}\right)= \sum_{L\geq 1} \big(g^2\big)^L\left(
  \frac{\gamma^{(L)}}{8}\frac{1}{(\epsilon L)^2}\left(1- \eps  L\log\left(\frac{s_1s_2}{\mu^2}\right)\right) - \frac{\gamma_{HgH}^{(L)}}{2\epsilon L}\right),
\end{equation}
where $\mu$ is the renormalization scale, and $\gamma^{(L)}$ is the coefficient of $(g^2)^L$ in the cusp anomalous dimension
defined below (\ref{Mhighernergy2}); $\gamma_{HgH}$, discussed below, is interpreted physically as the SCET collinear anomalous dimension of one massless field, plus that of two heavy fields. This Z-factor simultaneously renormalizes the IR divergences of both correlators, because the structure of the IR divergences arises only from the configuration of the external lines and not from the cusp point. Likewise UV divergences only come from the cusp and the operator insertion and not from the external scalar. 
The renormalization condition is then given by 
\begin{equation} \label{Renormalization_Condition}
 Z_{\rm IR}^{-1} \, \mathbf{Z}^{-1} \langle 0| \, \vec{W} | \Phi(p_3) \rangle={\rm finite} \, .
\end{equation}
Given the known cusp anomalous dimension (\ref{Z_cusp}), this equation allows us to determine the IR Z-factor and then the missing pieces in $\mathbf{Z}^{-1}$. 

\subsection{One-loop calculation}

\begin{figure}[t]

 \begin{minipage}{.3\textwidth} 
     \centering
      \begin{tikzpicture}[scale=1.7]
      
	  \coordinate (cusp) at (0,0) {};
	  \coordinate (InW) at (-1,-1) {};
	  \coordinate (OutW) at (1,-1) {};
	  \coordinate (InPhi) at (0,-1) {};
	  \coordinate (le) at (-0.5,-0.5) {};
	  \coordinate (ri) at (+0.5,-0.5) {};
	  
	  \draw[Wilson] (InW)  --  node[pos=0, anchor = north] {$v_1$} (le) -- (cusp) -- (ri) -- node[pos=1, anchor = north] {$v_2$} (OutW) {};
	  \draw[Scalar] (InPhi) -- node[pos=0, anchor = north] {$\Phi(p_3)$} (cusp) {};
	  
      \end{tikzpicture}
  \end{minipage}     
  \hfill  
 \begin{minipage}{.3\textwidth} 
     \centering
        \begin{tikzpicture}[scale=1.7]

	  \coordinate (cusp) at (0,0) {};
	  \coordinate (InW) at (-1,-1) {};
	  \coordinate (OutW) at (1,-1) {};
	  \coordinate (InPhi) at (0,-1) {};
	  \coordinate (le) at (-0.7,-0.7) {};
	  \coordinate (ri) at (+0.7,-0.7) {};
	  \coordinate (v) at (0,-0.7) {};
	  
	  \draw[Wilson] (InW)  --  node[pos=0, anchor = north] {$v_1$} (le) -- (cusp)  -- (ri) -- node[pos=1, anchor = north] {$v_2$} (OutW) {};
	  \draw[Scalar] (InPhi) -- node[pos=0, anchor = north] {$\Phi(p_3)$} (v) -- (cusp) {};
	  \draw[Gluon] (v) -- (le) {};
	  
      \end{tikzpicture}
     \end{minipage} 
     \hfill    
  \begin{minipage}{.3\textwidth} 
     \centering
        \begin{tikzpicture}[scale=1.7]

	  \coordinate (cusp) at (0,0) {};
	  \coordinate (InW) at (-1,-1) {};
	  \coordinate (OutW) at (1,-1) {};
	  \coordinate (InPhi) at (0,-1) {};
	  \coordinate (le) at (-0.7,-0.7) {};
	  \coordinate (ri) at (+0.7,-0.7) {};
	  \coordinate (v) at (0,-0.7) {};

	  \draw[Wilson] (InW)  --  node[pos=0, anchor = north] {$v_1$} (le) -- (cusp)  -- (ri) -- node[pos=1, anchor = north] {$v_2$} (OutW) {};
	  \draw[Scalar] (InPhi) -- node[pos=0, anchor = north] {$\Phi(p_3)$} (v) -- (cusp) {};
	  \draw[Gluon] (v) -- (le) {};
	  \draw[Gluon] (ri) -- (v) {};
	  
	  \filldraw [black] (v) circle [radius=0.7pt];
	  
      \end{tikzpicture}
  \end{minipage} 
\caption{Sample diagrams for the correlator $\langle 0| \, W_{\textrm{cusp},\Phi} |  \Phi(p_3) \rangle$. Double, curly and dashed lines represent Wilson lines, gluons and scalars, respectively}
\label{fig:Sample_Diagrams_W_Phi}
\end{figure}

\begin{figure}[t]

 \begin{minipage}{.3\textwidth} 
     \centering
      \begin{tikzpicture}[scale=1.7]

	  \coordinate (cusp) at (0,0) {};
	  \coordinate (InW) at (-1,-1) {};
	  \coordinate (OutW) at (1,-1) {};
	  \coordinate (InPhi) at (0,-1) {};
	  \coordinate (le) at (-0.45,-0.45) {};
	  \coordinate (ri) at (+0.45,-0.45) {};
	  
	  \draw[Wilson] (InW)  --  node[pos=0, anchor = north] {$v_1$} (le) -- (cusp) -- (ri) -- node[pos=1, anchor = north] {$v_2$} (OutW) {};
	  \draw[Scalar] (InPhi) -- node[pos=0, anchor = north] {$\Phi(p_3)$} (le) {};
	  
      \end{tikzpicture}
  \end{minipage}     
  \hfill  
 \begin{minipage}{.3\textwidth} 
     \centering
        \begin{tikzpicture}[scale=1.7]

	  \coordinate (cusp) at (0,0) {};
	  \coordinate (InW) at (-1,-1) {};
	  \coordinate (OutW) at (1,-1) {};
	  \coordinate (InPhi) at (0,-1) {};
	  \coordinate (le) at (-0.7,-0.7) {};
	  \coordinate (ri) at (+0.7,-0.7) {};
	  \coordinate (v) at (0,-0.7) {};
	  
	  \draw[Wilson] (InW)  --  node[pos=0, anchor = north] {$v_1$} (le) -- (cusp)  -- (ri) -- node[pos=1, anchor = north] {$v_2$} (OutW) {};
	  \draw[Scalar] (InPhi) -- node[pos=0, anchor = north] {$\Phi(p_3)$} (v) -- (ri) {};
	  \draw[Gluon] (v) -- (le) {};
	  
      \end{tikzpicture}
     \end{minipage} 
     \hfill    
  \begin{minipage}{.3\textwidth} 
     \centering
        \begin{tikzpicture}[scale=1.7]

	  \coordinate (cusp) at (0,0) {};
	  \coordinate (InW) at (-1,-1) {};
	  \coordinate (OutW) at (1,-1) {};
	  \coordinate (InPhi) at (-0.2,-1) {};
	  \coordinate (le) at (-0.7,-0.7) {};
	  \coordinate (le2) at (-0.25,-0.25) {};
	  \coordinate (ri) at (+0.7,-0.7) {};
	  \coordinate (v1) at (-0.2,-0.7) {};
	  \coordinate (v2) at (0.25,-0.7) {};

	  \draw[Wilson] (InW)  --  node[pos=0, anchor = north] {$v_1$} (le) -- (cusp)  -- (ri) -- node[pos=1, anchor = north] {$v_2$} (OutW) {};
	  \draw[Scalar] (InPhi) -- node[pos=0, anchor = north] {$\Phi(p_3)$} (v1) -- (v2) -- (ri) {};
	  \draw[Gluon] (v1) -- (le) {};
	  \draw[Gluon] (le2) -- (v2) {};
	  
	  
      \end{tikzpicture}
  \end{minipage} 
\caption{
Sample diagrams for the correlator $\langle 0| \, \partial{W}_{\rm cusp} |  \Phi(p_3) \rangle$. Double, curly and dashed lines represent Wilson lines, gluons and scalars, respectively}
\label{fig:Sample_Diagrams_W_Phi2}
\end{figure}
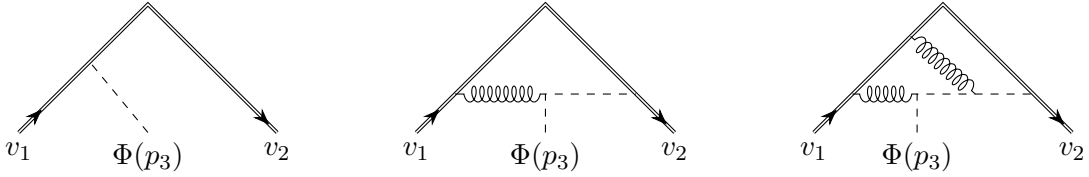

Let us now discuss in some detail the one-loop calculation.
 In Fig.~\ref{fig:Sample_Diagrams_W_Phi} and \ref{fig:Sample_Diagrams_W_Phi2} some sample Feynman diagrams are shown. 
We find the following result for the correlators, 
\begin{align}
 &\langle 0| \, \partial W_{{\rm cusp}}    | \Phi(p_3) \rangle =  \frac{(s_1+s_2)}{2 g_{\rm YM}}\langle 0| \, W_{\rm cusp} |  \Phi(p_3) \rangle   \label{Cor_Cusp} \\
 &\quad = \frac{(s_1+s_2)^2}{s_1 s_2} \left\{1+ g^2 \! \left(-\frac{1}{\epsilon^2} + \frac{1}{\epsilon} \left[\log\!\left(\frac{s_1 s_2}{\mu^2}\right)+\xi \, \log(x) \right] +\mathcal{O}\left(\epsilon^0\right) \right)  \! \right\} +\mathcal{O}\left(g^4 \right) \,, \nonumber  \\[12pt] 
&\langle 0| \, {W}_{{\rm cusp},\Phi} | \Phi(p_3) \rangle =  1+  g^2 \! \left(-\frac{1}{\epsilon^2}+  \frac{1}{\epsilon} \left[ \log\!\left(\frac{s_1 s_2}{\mu^2}\right)-2 \right] +\mathcal{O}\left(\epsilon^0\right)\right) +\mathcal{O}\left(g^4 \right) \, . \label{Cor_OP}
\end{align}
The IR Z-factor (\ref{Log_ZIR}) is determined by the first component of (\ref{Renormalization_Condition}),
with the result
\begin{align}
\gamma^{(1)}= 8, \qquad \gamma_{HgH}^{(1)}= 0\, . 
\end{align}
The first matches the expansion of the cusp anomalous dimension $\gamma(g^2)=8g^2-16g^4\zeta_2+\mathcal{O}(g^6)$ stated earlier.

The second component of (\ref{Renormalization_Condition}) then allows us to calculate the remaining pieces of $\mathbf{Z}^{-1}$ (or, equivalently $\mathbf{Z}$). 
We can deduce that at the one-loop level, the lower off-diagonal element in $\mathbf{Z}^{-1}$ has to vanish. This is seen as follows.
The tree level contribution of (\ref{Cor_Cusp}) to the second component in (\ref{Renormalization_Condition})  has the form 
$[g^2 Z_{\textrm{mix}}^{(1)} (s_1+s_2)^2]/[\epsilon s_1 s_2]$.  However there is no one loop contribution from (\ref{Cor_OP}) of this form and $Z_{\textrm{mix}}^{(1)}$ can not depend on $s_1$ and $s_2$, hence we have $Z_{\textrm{mix}}^{(1)}=0$. For the diagonal element we find $Z_{\textrm{cusp},\Phi}^{(1)}=-2$.

For the general case of a renormalization matrix, the corresponding anomalous dimension  matrix $\mathbf{\Gamma}$ is given by
\begin{align}
  \mathbf{\Gamma}= \mathbf{Z}^{-1} \frac{\d \mathbf{Z}}{\d \log \left(\mu \right)} \, , \qquad   \text{with} \quad \frac{\d g^2}{\d \log(\mu)}=-2 \epsilon g^2 \, .
\end{align}
Here we used that the $\beta$-function vanishes in $\mathcal{N}=4$ sYM theory. In our case the renormalization matrix is diagonal, and therefore the matrix inversion is trivial.
In this way, adding the engineering dimension of the scalar insertion, we obtain
\begin{align}
\Gamma_{{\rm cusp},\Phi} = 1+ 4 g^2 + \mathcal{O}(g^4) \,.
\end{align}
This is in agreement with our conjectured relation (\ref{mainconjecture}) (and also with the computation of the same anomalous dimension
in \cite{Alday:2007he}).

\subsection{Extension to two loops and discussion}

We proceeded to perform the calculation to two loops.
Some details of this calculation, and in particular the computation of the necessary Feynman integrals, can be found in Appendix \ref{app_soft_current}.
Proceeding as in the one-loop case and subtracting the known ultraviolet divergences of $W_{{\rm cusp}}$, we find for the IR renormalization coefficients
\begin{align}
\gamma^{(2)}= -16\zeta_2, \qquad \gamma_{HgH}^{(2)}= -2\zeta_3\,. \label{IR 2loops}
\end{align}
The cusp anomalous dimension matches the expected result, but it is also instructive to
compare the value of the second, single-logarithmic infrared divergence.
According to ref.~\cite{Becher:2009kw}, it should be the sum of a collinear anomalous dimension for massless particles,
plus the infrared anomalous dimensions for the two massive fundamental Wilson lines (or equivalently heavy particles):
\be
 \gamma_{HgH} = \gamma_g + 2 \gamma_{H,{\rm fund}}\,.
\ee
The known two-loop collinear anomalous dimension for any parton in this theory is $\gamma_g=2g^4\zeta_3$,
which matches with the maximal transcendental part
of the QCD result for either a gluon or an adjoint quark (see \cite{Dixon:2017nat}).
The massive case $\gamma_{H,{\rm fund}}$ was not calculated, to our knowledge,
however we can take the maximal transcendental part of the result for an adjoint QCD quark as given in \cite{Becher:2009kw}:
$\gamma_{H,{\rm adjoint}}= -4g^4\zeta_3\equiv 2\gamma_{H,{\rm fund}}$.
These indeed sum up to the value (\ref{IR 2loops}), which we conclude is
consistent with the principle of maximal transcendentality and the known QCD values.

Proceeding to the UV renormalization, we find that up to two loops, $\mathbf{Z}^{-1}$ remains diagonal,
and gives the anomalous dimension
\be
 \Gamma_{{\rm cusp},\Phi} = 1+ 4g^2 + g^4 \left[-\frac{1}{\xi } \left(4 \zeta _2 H_1+\frac{H_1^3}{6}\right)-8 \xi H_1 +2 H_1^2 +16 \left(\zeta _2+1\right)\right] + \mathcal{O}(g^6),
\ee
where $H$ are harmonic polylogarithms with argument $1-x^2$.
This is in perfect agreement with the conjectured relation (\ref{mainconjecture}) with the Regge trajectory given in eq.~(\ref{j1_threeloops}).

This two-loop calculation provides non-trivial evidence in favor of our proposed relation (\ref{mainconjecture}),
which identifies the subleading Regge trajectory $j_1$ with the scaling dimension $\Gamma_{{\rm cusp},\Phi}$ of a Wilson loop operator.
We note that the Regge trajectory $j_1$ is known (from the scattering amplitude calculation) to one more order, three-loops.
It would be interesting to verify that eq.~(\ref{mainconjecture}) also holds at that order, perhaps using integrability or other methods.

\section{Conclusion and outlook}  
\label{sec_conclusions}

In this paper we studied a massive four-particle scattering amplitude in $\cN=4$ super Yang-Mills.
Starting from the three-loop result obtained in \cite{Caron-Huot:2014lda}, we initiated a systematic analysis of this amplitude as a function of all kinematic invariants. While the full amplitude
involves complicated multiple polylogarithms that depend on two variables $s/m^2$ and $t/m^2$, it simplifies considerably in various physically interesting limits. In Appendix \ref{app_expansions}, we explain in detail how we obtained expansions from the differential equations of \cite{Caron-Huot:2014lda}, and
we provide the detailed three-loop formulas for the asymptotic expansions described in section \ref{sec_introduction_amplitudes_limits} in an ancillary file.

With the results of the asymptotic expansions at hand, we focused on the Regge limit.
Using the fact that the amplitude is governed by a hidden symmetry, dual conformal symmetry,
we derived a partial wave expansion which incorporates this additional information.
It was known that the leading term in the Regge limit exponentiates, with the exponent given by the anomalous dimension of a Wilson loop with a cusp.
In this paper, we explored subleading, power-suppressed terms.
Surprisingly, we found that, in this expansion, the first power suppressed term also exponentiates!
We computed the Regge exponent $j_1$ to three loops, cf. Eq. (\ref{j1_threeloops}).

Moreover, by using the symmetry to map the Regge kinematics to a soft expansion, cf. section \ref{sec:fromReggetoWilson},
we argued that the relevant operator controlling the subleading Regge limit should be a cusped Wilson loop with a scalar insertion at the cusp.
We verified this proposal to two loops in perturbation theory.
In order to do so, we performed a soft current calculation for massive quarks, to two loops, and found a perfect match.
The details of the calculation of the Feynman integrals are presented in Appendix \ref{app_soft_current}.

While we found that all integrals needed to compute the divergent part of the two-loop correlators required only multiple polylogarithms. 
It is interesting to mention that some of the finite integrals involve elliptic polylogarithms.

We briefly discuss some interesting questions for future work.
Our asymptotic expansions provide a wealth of data to explore higher order terms in the power expansion.
At sub-subleading power in the Regge limit ($1/s^2$),
we compared the expansion (\ref{subsubleading}) with an ansatz with two power laws.
We find that such an ansatz is inconsistent with angle-independent one-loop exponents, a property which would be expected in the perturbative expansion.
However, it is possible to write a consistent ansatz with three exponents.
Such an ansatz could be tested once higher-loop results for the scattering amplitude become available.

It would be interesting to derive a systematic expansion using heavy quark effective theory as sketched around eq.~(\ref{HQET}),
exploiting the setup of section \ref{sec:fromReggetoWilson} where the problem is mapped to a massless limit $m_3\to 0$.
The leading order terms correspond to the cusped Wilson loop, while we 
proposed that at first order one needs to consider a scalar insertion into the Wilson loop.
However, in general, the Lagrangian of this effective theory 
contains an infinite series of irrelevant operators suppressed by the heavy mass,
giving corrections to the Wilson lines, analogous to those considered in \cite{Laenen:2010uz,Luna:2016idw,Penin:2016wiw,Feige:2017zci}.
At higher orders in the power expansion, these may cause triangular mixing between
the operator (\ref{HQET}) and higher-dimensional ones. This could prevent the amplitude from being a sum of pure power laws.
It would thus be very interesting to elucidate the structure at higher powers,
and ultimately to translate these findings to the usual null Wilson lines approach to the Regge limit.

Once the operators are identified, a separate question consists in computing their anomalous dimensions.  
The cusp anomalous $\Gamma_{\rm cusp}$ is known to be governed by an integrable system \cite{Correa:2012hh,Drukker:2012de}, which was simplified in \cite{Gromov:2013pga}.
It would thus be interesting to see if the first subleading trajectory, given to three loops in eq.~(\ref{j1_threeloops}), can be reproduced quantitatively by extending the methods used in \cite{Gromov:2015dfa}.
We also wish to point out that though the AdS/CFT correspondence, it may be possible to study the anomalous dimensions at strong coupling \cite{Drukker:1999zq}.

\section*{Acknowledgements}
This research was supported in part by the National Science Foundation under Grant No. NSF PHY11-25915.
R.B. and J.M.H. are supported in part by the PRISMA Cluster of Excellence at Mainz university. This project has received funding from the European Research Council (ERC) under the European Union's Horizon 2020 research and innovation programme (grant agreement No 725110), `Novel structures in scattering amplitudes'.
R.B. is supported in part by a PhD fellowship of the Research Training Group 1581 `Symmetry breaking in fundamental interactions' of the DFG.

\appendix

\section{The total cross-section to three loops}  
\label{sec_cross_section}

Here we discuss the total cross-section $\sigma_{\rm tot} =  \sigma_{Y\bar Y\to X}$, see eq. (\ref{totalcrossPS}), 
up to three-loop order, using the results of \cite{Caron-Huot:2014lda}.
We will observe an interesting property about its high-energy behavior that motivates the all-loop prediction (\ref{sigmatot}).

In \cite{Caron-Huot:2014lda} the integrals contributing to the three- loop amplitude were computed. They fulfill a differential equation with trivial boundary conditions at $s=t=0$.  To study the forward limit $t=0$ we thus only need to integrate with respect to $s$, or equivalently
$u=\frac{4m^2}{-s}$. The differential equations have logarithmic-type singularities
at $u=0,-1$, and a square-root type singularity at $u=-1$.
Upon switching to the variable $x=\frac{\bu-1}{\bu+1}$, the alphabet becomes
$\{\log(x)$, $\log(1+x)$, $\log(1-x) \}$.
From this it follows that the limit of $M$ can be written in terms of harmonic
polylogarithms \cite{Remiddi:1999ew,Maitre:2005uu} with argument $x$.
Our result takes the form
\begin{align} \label{forwardlimit_oneloop}
 \lim_{t\to 0} \frac{m^2}{-t} M^{(1)} =& 2+ \frac{1+x}{1-x} \log (x) \,, \\
 \lim_{t\to 0} \frac{m^2}{-t} M^{(2)} =& -24\Li_3(-x)+16\log (x)\Li_2(-x)+4\log^2(x)\log(1+x)-18\zeta_3-4\zeta_2\log(x) \,.
 \end{align}
The two loop result originates solely from the horizontal ladder, $s^2t \GG{1,1,1,0}{1,0,1,1}{1}$, because the other integral is explicitly proportional to $st^2$.
For the same reason, at three-loops only two integrals from \cite{Caron-Huot:2014lda} contribute to the forward limit.
\begin{align}
\label{forwardlimit_threeloop}
 \lim_{t\to 0} \frac{m^2}{-t} M^{(3)} =&
 \frac{1-x}{1+x} \Big[ -96H_{0,-1,-1,0,0}(x)-32H_{0,-1,0,-1,0}+80 H_{0,-1,0,0,0}
 +32H_{0,-1,1,0,0}
 \nonumber \\
 &
 +128H_{0,0,-1,-1,0}-16H_{0,0,1,0,0}
+32H_{0,1,-1,0,0}+96H_{0,1,0,-1,0}
\nonumber\\
 &
 -64H_{0,1,0,0,0}+ \zeta_{2} \left( -16 H_{0,-1,0} + 64 H_{0,0,-1} + 48 H_{0,1,0}+24 \zeta_{3}  \right)\nonumber\\
 &
  +\zeta_{3} (120 H_{0,1}  - 8 H_{0,-1} )+28 \zeta_{4} H_{0} + 70 \zeta_{5} \Big]  
 \nonumber\\& -
 64H_{0,0,-1,0,0}+64H_{0,-1,0,0,0} -48H_{0,1,0,0,0}+48H_{0,0,1,0,0} +32 \zeta_4 H_0 + 120 \zeta_5 \,.
\end{align}
The first term originates from the ladder while the last line originates from the tennis court diagram.
Here $H$ are harmonic polylogarithms of argument $x$, which we
omitted for brevity.  Note that all formulas above are symmetric under $x \to 1/x$, as may be verified
by using identities between the harmonic polylogarithms for different arguments \cite{Remiddi:1999ew,Maitre:2005uu}.

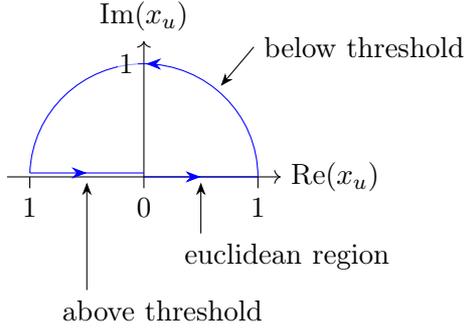
\begin{figure}[t]
\centering

\begin{tikzpicture}[scale=1.5]

\draw[axes_style] (-1.2,0) -- (1.2,0) node[right] {$\text{Re}(x_u)$};
\draw[axes_style] (0,0) -- (0,1.2) node[above] {$\text{Im}(x_u)$};


\draw[path_style] (0,0) -- (1,0);
\draw[path_style] (1,0) arc [start angle=0, end angle=178, radius=1];
\draw[path_style] (178:1) -- ++(1,0);

\draw[tick_style] (0,0) -- (0,-0.1) node[below] {0};
\draw[tick_style] (1,0) -- (1,-0.1) node[below] {1};
\draw[tick_style] (-1,0) -- (-1,-0.1) node[below] {1};
\node at (-0.05,1) [left, fill=white, opacity=0.6] {~};
\node at (0,1) [left] {1};

\draw[<-,pointer_style] (50:1.05) -- (50:1.5) node[right] {below threshold};
\draw[<-,pointer_style] (0.5,-0.05) -- (0.5,-0.5) node[anchor= 165] {euclidean region};
\draw[<-,pointer_style] (-0.5,-0.05) -- (-0.5,-1) node[anchor= 165] {above threshold};

\end{tikzpicture}
\caption{Analytic continuation in the forward limit regime, from $s<0$ (Euclidean region), to $0<s<4 m^2$ (below threshold), to $s>4 m^2$ (above threshold).}
\label{fig:analytic_continuation}. 
\end{figure}

The physical region above the threshold $s>4m^2$, where the inelastic cross-section (\ref{totalcrossPS}) is nonzero, correspond to $-1<x<0$. This region can be reached by analytic continuation. In total we can identify three regions relevant for the analytic continuation: the Euclidean region, below the threshold and above the threshold.
In the Euclidean region and below the threshold the amplitude is real. Above the threshold, however, the amplitude has a branch cut and following Feynman's $i0$ prescription we have to evaluate the harmonic polylogarithms slightly above the real axis. The analytic continuation is illustrated in Fig.~\ref{fig:analytic_continuation}.
By calculating the imaginary part of the amplitude we find for the total cross-section
\begin{align}\label{sigmatot_threeloops}
 \sigma_{Y\bar{Y}\rightarrow X}= \frac{2\pi g_{\rm YM}^2}{m^2} \left[ g^2 X_1  + g^4 X_2 + g^6 X_3 +   \mathcal{O}\big(g^8 \big) \right]\,,
 \end{align}
 where
 \begin{align}
 X_1 =&  \frac{1+x}{1-x} \,,\\
  X_2 =&    16 {\rm Li}_2(-x)+8 \log (-x) \log (x+1)-\frac{2 \pi ^2}{3}  \,,\\
  X_3 =&  -48 H_{-3,0}(-x)+64 H_{3,0}(-x)+48 H_{-2,0,0}(-x)-64 H_{2,0,0}(-x)\nonumber \\
  & -48 \zeta_2 H_{-2}(-x)+64 \zeta_2 H_2(-x)+32 
   \zeta_4 \nonumber\\ & + \frac{1-x}{1+x} \Big[ 16 H_{-3,0}(-x)+96 H_{-2,2}(-x)   -32 H_{2,2}(-x) +128 H_{3,1}(-x)\nonumber\\ &+64 H_{-2,0,0}(-x)+32 H_{-2,1,0}(-x)+32
   H_{2,-1,0}(-x)-80 H_{2,0,0}(-x)\nonumber\\ &-96 H_{2,1,0}(-x)-112 \zeta_2 H_{-2}(-x)+96 \zeta_2 H_2(-x)+28 \zeta_4  \Big] \,.
\end{align}
Here we have chosen a form that is manifestly real-valued for $-1<x<0$.

The cross-section (\ref{sigmatot_threeloops}) can be seen to approach a constant in the
high-energy limit $x\to 0$:
\begin{align}
X_1 \to 1 \,,\qquad X_2 \to  -\frac{2 \pi^2}{3} \,,\qquad X_3 \to \frac{2 \pi^4}{3} \,. \label{X1X2X3}
\end{align}
Remarkably, this agrees precisely with the perturbative expansion of the Bremsstrahlung function (\ref{Bremsstrahlungexact})!
This is not a coincidence. To see this, we use the leading Regge behavior of the amplitude
at large $s$ and fixed $t$
\be
 \lim_{s\to\infty} M\left(\frac{4m^2}{-s},\frac{4m^2}{-s}\right)= \tilde{r}_0(t) (-s-i0)^{1+j_0(t)} +\mathcal{O}(1/s)\,,
\ee
where $\tilde{r}_0$ and $j_0+1$ are given in (\ref{LO_trajectory1}).
Because of the mass gap of the $W$ bosons, loop corrections to the \emph{amplitude} $A\propto Ms/t$ must be real and analytic around $t=0$,
which implies that the coefficient of $1/t$ is tree-level exact. Thus the loop
corrections to these parameters must vanish at the origin: $\tilde{r}_0(0)=1$ and $j_0(0)=-1$.
Furthermore the coefficient $\tilde{r}_0$ is real, so the imaginary part originates from the trajectory.
Thus
\begin{align}
\lim_{t\to 0}  \lim_{s\to \infty}  \frac{1}{-t} \,{\rm Im} \,M(s,t) = \pi \frac{d}{dt}j_0(t)\big|_{t=0}. \label{unusual}
\end{align}
Equation (\ref{unusual}) is a bit unusual since the cross-section
involves the slope of the Regge trajectory at $t=0$ rather than the intercept, as is more usual.
This happens here because the intercept precisely vanishes.
Using the relation (\ref{Bremsstrahlungexact}) expressing the slope at $t=0$ in terms of the (exactly known) Bremsstrahlung function
gives the prediction (\ref{sigmatot}) for the total cross-section, generalizing (\ref{X1X2X3}) to all orders.

\section{Regge expansion using dual conformal partial waves}
\label{app_partial_waves}

As discussed in section \ref{sec_subleading_regge}, the enhanced symmetry of the amplitude we look at
makes it possible to efficiently organize the Regge limit.
In this section we derive this improved expansion which exploits dual conformal symmetry.

\begin{figure}[t]
\centering

    \begin{tikzpicture}[scale=1.2]

	  \coordinate (a11) at (0,0) {};
	  \coordinate (a12) at (1,0) {};
	  \coordinate (a13) at (2,0) {};
	  \coordinate (a21) at (0,-1) {};
	  \coordinate (a22) at (1,-1) {};
	  \coordinate (a23) at (2,-1) {};
	 
	  
	  \draw[ultra thick] (a11)  --  (a13) --  (a23) --   (a21) --  (a11) {};
	  \draw[dashed] (a12) -- (a22);
	  \draw[dashed] (a11) --  ++(-0.5,0.5);
	  \draw[dashed] (a13) --  ++(+0.5,0.5);
	  \draw[dashed] (a21) --  ++(-0.5,-0.5);
	  \draw[dashed] (a23) --  ++(+0.5,-0.5);

	  \node at (-0.5,-0.5) {$Y_1$};
	  \node at (1,0.5) {$Y_2$};
	  \node at (2.5,-0.5) {$Y_3$};
	  \node at (1,-1.5) {$Y_4$};
	  \node at (0.5,-0.5) {$X_1$};
	  \node at (1.5,-0.5) {$X_2$};
	  
    \end{tikzpicture}
\caption{Visualization of the dual coordinates defined in (\ref{XX XY})-(\ref{Y_embed}). Solid and dashed lines represent massive and massless particles, respectively}
\label{fig:Dual_Coordinates}
\end{figure}
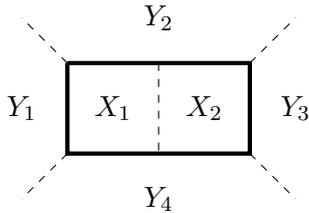

It will be helpful to use the embedding formalism,
which realizes Minkowski space as the null cone in R${}^{4,2}$.
The region momenta associated to each (planar) loop are represented as a (projective)  6-vector
\be
 X^A = (\vec{x},\tfrac{x^2}{2\mu}-\tfrac{\mu}{2} \,\big|\, x^0, \tfrac{x^2}{2\mu}+\tfrac{\mu}{2})\,,
\ee
where the first four components (before the vertical line) are spacelike and the last two are timelike.
Here $\mu$ is an arbitrary scale and $x^2=\vec{x}^2-(x^0)^2$.
In the presence of internal masses we also need to consider timelike dual coordinates for the external regions (see fig.~\ref{fig:Dual_Coordinates}), satisfying $Y_i^2=-m^2$:
\be \label{Yi}
 Y_i^A = (\vec{y},\tfrac{y^2+m^2}{2\mu}-\tfrac{\mu}{2}\,\big|\, y^0, \tfrac{y^2+m^2}{2\mu}+\tfrac{\mu}{2}).
\ee
These definitions ensure that six-dimensional dot products give massless and massive momentum space propagators:
\be
-2X_i\cdot X_j =  (x_i-x_j)^2\,,\qquad -2X_i\cdot Y_j =  (x_i-y_j)^2+m^2\,. \label{XX XY}
\ee
This notation is helpful because the dual conformal symmetry SO(4,2) acts linearly as rotations of these 6-vectors.
The external momenta of our planar four-particle scattering problem are encoded
in the differences between the above points $Y_i^A$ as: $y_{i+1}^\mu-y_{i}^\mu=p_i^\mu$.
For definiteness let us begin by assuming kinematics with a timelike $t$-channel, with $0<t<4m^2$ where $t=-(p_2+p_3)^2$.
We can use Lorentz invariance to go to the rest frame of $p_2+p_3$
and use translation invariance in $y$-space to set $\vec{y_2}=\vec{y_4}=0$.
Furthermore the energies of $p_1,p_2$ must be equal and opposite, which allows to set $y_1^0=y_3^0=0$.
In this frame the dual coordinates reduce to
\be
Y_1 = \left(\begin{array}{c} -\vec{p}_1 \\  \frac{t}{2\alpha} \\\hline  0\\ \tfrac{2m^2}{\alpha} \end{array}\right),\quad
Y_2 = \left(\begin{array}{c} \vec{0} \\ 0 \\\hline \tfrac12\sqrt{t} \\ \tfrac\alpha2 \end{array}\right),\quad
Y_3 = \left(\begin{array}{c} \vec{p}_2 \\ \frac{t}{2\alpha} \\\hline 0 \\ \tfrac{2m^2}{\alpha} \end{array}\right),\quad
Y_4 = \left(\begin{array}{c} \vec{0} \\ 0 \\\hline -\tfrac12\sqrt{t} \\ \tfrac\alpha2 \end{array}\right), \label{Y_embed}
\ee
where $\alpha=\sqrt{4m^2-t}$ and in addition we have chosen $\mu=\frac{\alpha}{2}$ in order to set to zero the last spacelike component of $Y_2$ and $Y_4$.

The SO(4) symmetry is apparent in this frame: these are simply the rotations of the first four components,
which preserve the two dual coordinates $Y_2$ and $Y_4$ bounding the $t$-channel.
This contains the usual SO(3) rotations of a pair of particles in its rest frame, with three additional generators
which are related, in the nonrelativisitc limit, to the flow generated by the Laplace-Runge-Lenz vector of the hydrogen atom \cite{Caron-Huot:2014gia}.

From eq.~(\ref{Y_embed}) we see that the dependence on the Mandelstam variable $s=-(p_1+p_2)^2$ is encoded in the SO(4)-invariant angle between the first four components of $Y_1$ and those of $Y_3$. From a short computation:
\be
 \cos\theta\equiv \frac{-\vec{p_1}{\cdot}\vec{p_2}+\frac{t^2}{4\alpha^2}}{\vec{p_1}^2+\frac{t^2}{4\alpha^2}}= 1+\frac{2s}{t} - \frac{s}{2m^2}
 \label{SO4_angle}
\ee
where we have simplified using that $\vec{p}_1^2=\vec{p}_2^2=\frac{t}{4}$ and $-\vec{p}_1\cdot \vec{p}_2=\frac{t}{4}+\frac{s}{2}$.
As noted in the text, this angle differs from the usual scattering angle between the
two external massless photons (see eq.~(\ref{SO(3)})) by the $-s/(2m^2)$ term.
It is real, for example, in the $t$-channel region where $0<t<4m^2$ and $-t<s<0$.

We now derive the corresponding partial wave expansion, starting from the case where the angle is real and then analytically continuing.
The idea is express the dependence on $s$ in terms of SO(4) spherical harmonics; for each SO(4) spin $j$
these sum up to a Chebyshev polynomial (the SO(4) analog of the Legendre polynomials):
\be
A= \sum_{j=0}^\infty
c_j(t) P'_j(\cos\theta), \qquad P'_j(\cos\theta) = \frac{\sin((j+1)\theta)}{(j+1)\sin(\theta)}\,. \label{SO4_poly}
\ee
Using the relation similar to (\ref{ampYYbar}) between the $Y\bar{Y}\to \bar{Y}Y$ amplitude and the stripped matrix element $M$,
\be
A= \frac{t}{s}M(s,t)=\frac{1-t/4m^2}{\sin^2(\theta/2)} M(s,t)
\ee
and absorbing $s$-independent factors into the coefficients $c_j(t)$, we can rewrite this as an expansion for $M$:
\be
\frac{1+Y}{1-Y} M(s,t) = \sum_{j=0}^\infty c'_j(t) (Y^{j+1}-Y^{-j-1})\,,\qquad Y\equiv e^{i\theta} = \frac{\buv-\bv}{\buv+\bv}\,. \label{SO4_sum}
\ee
We note that, as a mathematical statement about bases of functions,
one could equally well apply this decomposition to $M$ itself (or
to $M$ times times any function of the cross-ratios $u, v$).
The amplitude $A$ is singled out physically since its $t$-channel cuts have a Hilbert space interpretation in terms of intermediate
states acted upon by SO(4).\footnote{This can be checked from the numerators of the various contributions to $M$,
for example the $L$-loop $t$-channel ladder. The $s$-dependence of its coefficient $s t^L$, which is non-factorized
since it couples directly the two endpoints $Y_1,Y_3$ of the ladder, cancels in the amplitude we use.}

We now have an expansion valid for real angles.
To reach the Regge limit $Y\to 0$ where the angle is imaginary, we follow the standard procedure
and rewrite the sum as an integral using the Watson-Sommerfeld trick \cite{Donnachie:2002en,Collins:1977jy}:
\be
\frac{1+Y}{1-Y} M(s,t) = \int\limits_{-\eps-i\infty}^{-\eps+i\infty} \frac{i\,dj}{2\sin(\pi j)} c_j(t) (e^{-i\pi j}Y^{j+1}-e^{i\pi j}Y^{-j-1})\,,
\ee
where we have assumed that $0<\theta<\pi$ and the phases have been chosen such that the integrand vanishes at large imaginary $j$
(assuming that $c_j(t)$ is bounded).  The idea is that, deforming the contour to the right and picking up the residues of
$\sin(\pi j)$, this reproduces the sum (\ref{SO4_sum}).
But taking the Regge limit $Y\to 0$, a different contour deformation becomes appropriate.
The contour can still be closed to the right in the first term, but now to the left in the second term.

Two types of singularities arise: poles from the inverse sine factor, which add up to:
\be
 c_{-1}(t) +\sum_{k\geq 0} Y^{k+1}\left(c_{k}(t) + c_{-2-k}(t)\right)\,.
\ee
These however neatly cancel out, because for \emph{integer} spin $j$ the coefficients $c_{j}$ are odd under $j\to -j-2$.
Such a cancellation of kinematic poles
can be proved from the Froissard-Gribov inversion formula and
occurs generally for any SO(D) expansion; a detailed discussion in the SO(3) case can be found in refs.~\cite{Donnachie:2002en,Collins:1977jy}.
All that remain are the Regge poles of $c_k(t)$ from the second term,
which add up to the asymptotic expansion:
\be
\lim_{Y\to 0} \frac{1+Y}{1-Y} M = \sum_{n=0}^\infty r_n(t) Y^{-j_n(t)-1}\,,\label{SO4_final_appendix}
\ee
where we have defined the residues $r_n(t)=\frac{\pi e^{i\pi j_n(t)}}{\sin(\pi j_n(t))} {\rm Res}_{j=j_n(t)}\,c_j(t)$.
This formula is used in the main text to efficiently organize the Regge expansion $Y\sim \frac{1}{s}\to 0$.

\section{Method for expanding the three-loop amplitude of \cite{Caron-Huot:2014lda}}
\label{app_expansions}

\begin{figure}[t]
\centering

\begin{tikzpicture}[scale=0.8]
\draw[axes_style] (0,0) -- (5.5,0) node[right] {$u$} node[midway,above,align=center] {Regge limit \\ $t \rightarrow \infty$};
\draw[axes_style] (0,0) -- (0,5.5) node[above] {$v$} node[midway,left,align=center] {Regge limit \\ $s \rightarrow \infty$};

\draw[help_lines_style] (-0,5) -- (5,5) node[midway,below,align=center] {forward limit \\ $t \rightarrow 0$} ;
\draw[help_lines_style] (5,0) -- (5,5) node[midway,right,align=center] {forward limit \\ $s \rightarrow 0$};

\filldraw [black] (0,0) circle [radius=2pt]
		  (5,5) circle [radius=2pt];

\draw[tick_style] (0,-0.2) -- (0,0);
\draw[tick_style] (5,0) -- (5,-0.2) node[below] {$\infty$};
\draw[tick_style] (0,0) -- (-0.2,0);
\draw[tick_style] (0,5) -- (-0.2,5) node[left] {$\infty$};
\node at (0,0) [anchor = north east] {0};

\draw[<-,pointer_style] (5.15,5) -- (6,5) node[right, align=center] {soft limit \\ $s,t \rightarrow 0$};

\draw[<-,pointer_style] (-0.1,0.1) -- (-1,0.8) node[left, align=center] {high energy limit \\ $s,t \rightarrow \infty$};

\end{tikzpicture}
\caption{Different limits we consider in the $u-v$ plane. To derive expansions, first the boundary value for each limit
is obtained. Initially known in the soft limit, the boundary value  is transported along the edge of the diagram. }
\label{fig:boundary_transport}
\end{figure}
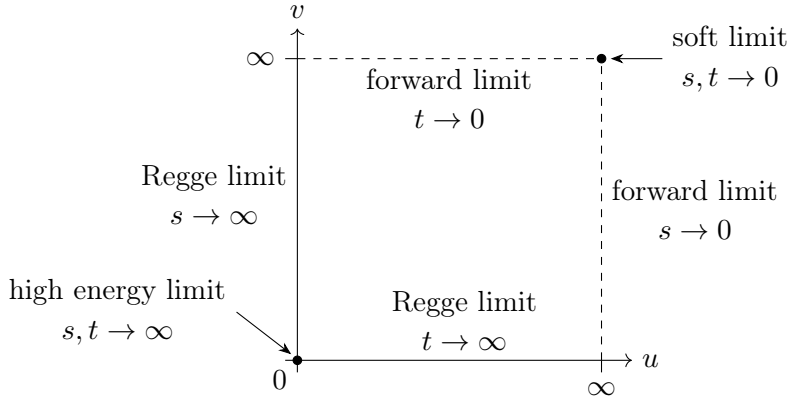

Here we explain how to express the amplitude in various limits, which in general
can contain logarithmic divergences.
In principle, we could use the analytic expressions for the master integrals
derived in ref. \cite{Caron-Huot:2014lda}, and expand them using properties of the iterated integrals
they were expressed in. We find it more convenient to obtain such expansions 
directly from differential equation for the master integrals that were derived in ref. \cite{Caron-Huot:2014lda}.

In order to do so, we use a well-known procedure for solving differential equations
in a limit, following closely the textbook \cite{Wasow}. Let $x$ be parameter that parametrizes
the expansion around $x=0$, and let $\mi{f}$ be the vector of master integrals.
As we will see, the solution for $\mi{f}$ takes the general form $P(x) x^{A_0} \mi{f}_0$, where $P(x)$ 
is a (matrix) polynomial in $x$; the matrix exponential $x^{A_{0}}$ contains possible logarithmic divergences,
and  $\mi{f}_{0}$ is the finite boundary value at $x=0$. Given possible powers of logarithms $\log(x)$,
one may also call $\mi{f}_{0}$ the `regularized' boundary value.

A technical point is related to obtaining such boundary values for all expansions that we are interested in.
The boundary value considered in ref. \cite{Caron-Huot:2014lda} is taken at $s,t \to 0$, see Fig. \ref{fig:boundary_transport}.
In order to obtain appropriate boundary values for other expansions, we first transport this value
to other regions, along appropriate paths. By `transporting' we mean solving the differential equation
along a given path. In principle one could choose any convenient path.
However, some choices are preferable over others.  In particular, one can often find paths
for which the one-parameter solution  is expressible in terms of a relatively simple class of 
functions, the harmonic polylogarithms. This is the case for the paths shown in Fig. \ref{fig:boundary_transport}.

As we will discuss in more detail in the following, special care is required when singular boundaries are
approached (corresponding to singularities of the differential equation). When several of such boundaries
intersect, it is important to clarify how the singular boundary is approached. In mathematical language,
one can perform a `blowup' that resolves singular intersections of boundaries.

As a non-trivial verification of our analytic continuation procedure, we verified that, upon returning to
the original point $s,t \to 0$ after going around the whole square in the positive quadrant shown in
Fig. \ref{fig:boundary_transport}, we recover the correct boundary value.

\subsection{Solving the differential equation in an expansion}

In this section we follow \cite{Wasow} closely.
Given a square $n$-th order matrix $\bar{A}(x)$, which is holomorphic on a connected open set $R \subset \mathbb{C}$, the differential equation $\mi{f}'(x)=\bar{A}(x) \mi{f}(x) $
has a unique solution on $R$, provided a boundary condition $\mi{f}(a)=\mi{f}_{BV} \, , \; a\in R $. Furthermore this solution is holomorphic on $R$. We are interested in the more special case where the matrix $\bar{A}(x)$ has a regular singular point $x_p \notin R$. Without loss of generality we choose $x_p=0$. Then the differential equation can be rewritten as
\begin{equation} \label{DE_Vec}
  x \mi{f}'(x)=A(x) \mi{f}(x) \, .
\end{equation}

As a first step in solving (\ref{DE_Vec}) we perform a transformation $\mi{f}(x)=P(x) \mi{g}(x)$ with a non-singular holomorphic matrix $P(x)$ to get
\begin{equation} \label{DE_Mat_Help}
 x \mi{g}'(x)=B(x) \mi{g}(x) \, .
\end{equation}
The new matrix $B(x)$ is determined by $P(x)$ and $A(x)$. 
Our aim is now to find a $P(x)$ such that $B(x)$ becomes as simple as possible, in order to solve (\ref{DE_Mat_Help}). 
As we will see, in practice, we can choose $B(x)$ and then calculate $P(x)$ using
\begin{equation*}
 xP'(x)=A(x)P(x)-P(x)B(x) \, .
\end{equation*}
Inserting the respective power series for $A(x)=\sum_{k\in\mathbb{N}_0} A_k x^k$ as well as for $B(x)$ and $P(x)$ in the differential equation above, we obtain, after equating the coefficients, the following recursion relation
\begin{align}
 A_0 P_0-P_0 B_0 &=0 \\
 (A_0-k  \mathbbm{1}) P_k -P_k B_0 &=- \sum_{j=0}^{k-1} (A_{k-j} P_j-P_j B_{k-j}) \, , \; k>0 \, . \label{Recursion}
\end{align}
At this point a subtleness arises: We are of course interested in an unique solution of the problem, but the matrix equation $A X- X B=0$ for given square matrices $A$ and $B$ can in principle has a non-trivial solution for the matrix $X$. One can show that the equation $A X- X B=0$  has such a  non-trivial solution $X \neq 0$ if and only if $A$ and $B$ have at least one common eigenvalue.

Equipped with this knowledge, we now choose a certain matrix $B(x)$. If $A$ in (\ref{DE_Vec}) is a constant matrix, then we do not need to simplify the problem any further. Therefore we choose as our starting point $B_0=A_0$ and $P_0=\mathbbm{1}$.

From the previous argument we know that if no pair of eigenvalues of the matrix $A_0$ differs by a positive integer, the $P_k$ are determined by ($\ref{Recursion}$).
In our case the matrix appearing in the differential equation for the master integrals is a lower triangular matrix with vanishing diagonal elements, hence all eigenvalues are zero and the former condition is trivially fulfilled.
We choose $B_k=0$ for $k>0$ in order to obtain a simple differential equation after the transformation. 
With this choice $B(x)=A_0$ the solution of (\ref{DE_Mat_Help}) is given by $x^{A_0}$. 
  Returning to the original problem, we find the asymptotic expansion of the solution of (\ref{DE_Vec})
\begin{equation}
  \mi{f}(x)=P(x) x^{A_0} \mi{f}_0=P(x) \exp[A_0 \log(x)]  \mi{f}_0 \, ,
\end{equation}
where $P(x)$ is calculated recursively from (\ref{Recursion}) using $P_0=\mathbbm{1}$, $B_0=A_0$ and $B_k=0$ for $k>0$.

We wish to make the following comments.
\begin{itemize}
 \item As was already mentioned earlier, the solution may have logarithmic divergences in the limit $x\to 0$.
If present, these are described by the matrix exponential $ \exp[A_0 \log(x)]$.
We call $\mi{f}_0$ the boundary value at $x=0$, even in such singular cases.
\item One can interpret the matrix $F(x) = P(x) x^{A_0}=P(x) \exp[A_0 \log(x)] $ as the fundamental system of solutions of the 
matrix differential equation $x \, F'(x) = A(x) F(x)$.
\end{itemize}

In the following subsections, we describe in more detail the procedure of obtaining the boundary values for the different expansions, and on the choice of variables for the latter.

\subsection{Soft expansion}

The soft or low energy limit describes the region where $|s|, |t| \ll  4m^2$. We perform the calculation in the Euclidean region $s,t<0$, but the result is valid in the entire region $|s|, |t| \ll  4m^2$, since the result is simply a polynominal in $s$ and $t$.

In order to derive the soft expansion, we can use our starting boundary value at $m \rightarrow \infty$ with $\mi{g}_{\rm start}=(1,0,...,0)$. 
For solving the differential equation we introduce the transformation
\begin{equation} \label{Trafo_soft}
 s=- \frac{4m^2(1+R) }{R} x^2 \, , \quad t=- 4m^2(1+R)x^2 \qquad \bigg( \Leftrightarrow u+v= \frac{1}{x^2} \, , \quad R=\frac{u}{v} \bigg) \,.
\end{equation}
with the ratio $R=u/v=s/t$ fixed. The differential equation is then solved for small $x$, as explained above. 

Applying this to the amplitude, we find, up to three loops,
\begin{equation}
\begin{aligned}
\frac{M-1}{st} \approx & -\frac{g^2}{6m^4} -\frac{s+t}{m^6} \left( \frac{g^2}{60}+\frac{g^4}{12}-\frac{g^6}{3}\right) \\ &-\frac{st}{m^8} \left(\frac{g^2}{840}+\frac{g^4}{180}\right) - \frac{s^2+t^2}{m^8}\left(\frac{g^2}{420}+ \frac{g^4}{45}-\frac{g^6}{24} \right)\,.
\end{aligned}
\end{equation}
As mentioned earlier, the $1/m^4$ term is one-loop exact.
In the perturbative expansion this feature is evident already pre-integration:
all higher-loop integrals appearing in the perturbative expansion
are explicitly proportional to at least $s^2t$ or $st^2$, e.g. see Fig.~7 of \cite{Bern:2007ct} for the five loop integrand.

It is noteworthy that all coefficients are rational multiples of $g^2=g_{YM}^2 N /(16\pi^2)$; no transcendental numbers such as $\zeta$ values appear.
Technically this can be traced to the fact that in ref. \cite{Caron-Huot:2014lda}
a uniform weight basis could be found in which all but one integral vanish in the low-energy limit.
It would be interesting to see if this remains the case at higher loop orders.

\subsection{Regge expansion}

Switching from the kinematic invariants $s$ and $t$ to the variables $u$ and $v$, the Regge limit is described by $v \gg u$. In our calculation we consider the limit $u \rightarrow 0$ with $v >0$. To transport the boundary value from our starting point at $(u,v)=(\infty,\infty)$ (this is the limit $m \rightarrow \infty$ in the Euclidean region) to our end point,  
we split the path in two straight line segments $\gamma_1$ and $\gamma_2$. The first path $\gamma_1=\{(u,v)=(-t,\infty)|t \in(- \infty,0]\}$ is parallel to the $u$-axis and ends on the $v$-axis, while the second path $\gamma_2=\{(u,v)=(0,-t)|t \in(- \infty,-\tilde{v}]\}$ is on the $v$-axis and ends at some $\tilde{v}>0$.

For the first segment $\gamma_1$ we take the limit $v\rightarrow \infty$ and substitute the variable $u$:
 \begin{equation} \label{Trafo_Regge}
  u=\frac{4x_u}{(1-x_u)^2} \quad \Leftrightarrow \quad x_u=\frac{\beta_u -1}{\beta_u+1} \, \qquad {\rm with} \; u \in [0,\infty) \, , x_u \in [0,1) \, .
 \end{equation}
 This leads to a differential equation for the master integrals $\mi{g}$ on the path $\gamma_1$ with the alphabet $\{\log(x_u)$, $\log(1 - x_u)$, $\log(1 + x_u)\}$\,,
 \begin{equation}
  \frac{\d \mi{g}^{\gamma_1}}{\d x_u}=A^{\gamma_1}(x_u) \mi{g}^{\gamma_1}=\left[ \lim_{v \rightarrow \infty} \frac{\partial A(x_u,v)}{\partial x_u}\right] \mi{g}^{\gamma_1} \, .
 \end{equation}
 Since $A^{\gamma_1}$ is a  lower triangular matrix 
 we obtain the solution recursively.
 It can be expressed in terms of harmonic polylogarithms and constants, which are determined by the boundary value at $x_u = 1$. 
For the path $\gamma_2$ we need the boundary value at $x_u=0$. Unlike at $x_u=1$ the master integrals exhibit logarithmic divergences at $x_u=0$. 
To extract the (regularized) boundary value, we use the asymptotic expansion of the differential equation.

The calculation for the second path $\gamma_2$ is identical to the previous one. In the limit $u \rightarrow 0$ and with the variable transformation (\ref{Trafo_Regge}) for the variable $v$ the differential equation can be solved on the path $\gamma_2$ in terms of harmonic polylogarithms. With the previous calculated boundary value at $x_u=0$ the integration constants are fixed. 
With the solution $\mi{g}^{\gamma_2}(x_v)$ on the second path as our boundary value we finally solve the differential equation in an asymptotic expansion with arbitrary $v > 0$ in $x_u$ near $x_u=0$ to obtain the master integrals in the Regge limit.

\subsection{High energy expansion}

\begin{figure}[t]
\begin{minipage}{0.4\textwidth}
\centering

\begin{tikzpicture}[scale=1.3]

\draw[axes_style] (0,0) -- (2.5,0) node[right] {$u$};
\draw[axes_style] (0,0) -- (0,2.5) node[above] {$v$};

\draw[tick_style] (0,-0.1) -- (0,0);
\draw[tick_style] (1.5,0) -- (1.5,-0.1) node[below] {$\delta$};
\draw[tick_style] (0,0) -- (-0.1,0);
\draw[tick_style] (0,1.5) -- (-0.1,1.5) node[left] {$\delta$};
\node at (0,0) [anchor = north east] {0};

\draw[ana_strct_style] (-1,1) -- (1,-1) node[very near end, right] {$\;v=-u$};
\draw[ana_strct_style] plot[smooth] file {Plots/fct1.table} node[anchor=210] {$u^2=4v$};
\draw[ana_strct_style] plot[smooth] file {Plots/fct2.table} node[anchor=210] {$v^2=4u$};

\draw[path_style] (1.5,0) -- (0,1.5) node[midway, right] {$\gamma_\delta$};

\node at (-0.8,2.6) {(a)};

\end{tikzpicture}
\end{minipage}
\hfill
\begin{minipage}{0.55\textwidth}
\centering

\begin{tikzpicture}[scale=1.3]

\draw[axes_style] (0,0) -- (3.7,0) node[right] {$t$};
\draw[axes_style] (0,0) -- (0,2.5) node[above] {$\delta$};

\draw[help_lines_style] (3.5,0) -- (3.5,2.5);

\draw[tick_style] (0,-0.1) -- (0,0);
\draw[tick_style] (0.875,0) -- (0.875,-0.1) node[below] {$\epsilon$};
\draw[tick_style] (2.625,0) -- (2.625,-0.1) node[below] {$1-\epsilon$};
\draw[tick_style] (3.5,0) -- (3.5,-0.1) node[below] {1};
\draw[tick_style] (0,0) -- (-0.1,0);
\draw[tick_style] (0,1.777) -- (-0.1,1.777) node[left] {$\frac{4 \, \epsilon}{(1-\epsilon)^2}$};
\node at (0,0) [anchor = north east] {0};

\draw[ana_strct_style] plot[smooth] file {Plots/Tfct1.table} node[above] {$u^2=4v$};
\draw[ana_strct_style] plot[smooth] file {Plots/Tfct2.table} node[above] {$v^2=4u$};

\draw[path_style] (0,1.777) -- (0.875,1.777) node[midway, above] {$\gamma_\epsilon^1$};
\draw[path_style] (0.875,1.777) -- (0.875,0) node[midway, right] {$\gamma_\epsilon^2$};
\draw[path_style] (0.875,0) -- (2.625,0) node[midway, above] {$\gamma_\epsilon^3$};
\draw[path_style] (2.625,0) -- (2.625,1.777) node[midway, left] {$\gamma_\epsilon^4$};
\draw[path_style] (2.625,1.777) -- (3.5,1.777) node[midway, above] {$\gamma_\epsilon^5$};

\node at (-0.8,3) {(b)};

\end{tikzpicture}
\end{minipage}
\caption{Left: Choice of path in the $u$-$v$-plane. Right: Choice of path in the $t$-$\delta$-plane.}
\label{fig:high_energy_path}
\end{figure}
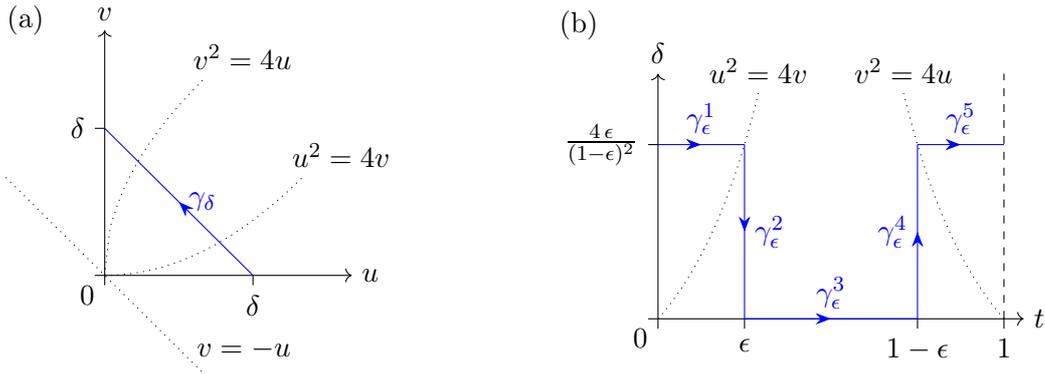

In the high energy limit we have $|s|,|t| \gg 4m^2 $. We will work in the Euclidean region. 
Obtaining the appropriate boundary value at $(u,v)=(0,0)$ requires some care, as we
discuss presently.
The subtlety originates from the singularity structure of the differential equation in the limit $u,v \rightarrow 0$. 
This can be immediately understood by inspecting the alphabet of the differential equation \cite{Caron-Huot:2014lda}, which
contains the letters $\{\log(u)$, $\log(v)$, $\log(u+v)$, $\log(u^2-4v)$, $\log(v^2-4v)\}$. It is sufficient to study these ``simple'' letters, because the other letters do not add more singularities in the vicinity of $(u,v)=(0,0)$.  The corresponding singular lines are shown in Fig. \ref{fig:high_energy_path}(a).
The fact that the latter intersect at $(u,v)=(0,0)$ implies that one has to specify how exactly this point is approached.
The problem of a potential ambiguity can be avoided by switching to appropriate variables that
resolve the way the singularity is approached.

The variable transformations can also be understood as choosing more sophisticated paths near the origin to connect the boundary values, which we obtain by approaching the origin $(u,v)=(0,0)$ on the $u$-axis or $v$-axis. These two boundary values are denoted by $\mi{g}_{BV}^u$ and $\mi{g}_{BV}^v$.
In Fig.~\ref{fig:high_energy_path} these paths are shown. The first transformation or path
\begin{equation} \label{Trafo_high_energy}
 \gamma_\delta(t)=(u,v)=\big(\delta (1-t),\delta t \big) \, ,\quad t\in [0,1] \, , \quad \delta>0
\end{equation}
resolves the ambiguity of the first three considered letters
\begin{equation*}
 \{\log(u),\log(v),\log(u+v) \} \longrightarrow \{\log(\delta),\log(t),\log(1-t) \} \, .
\end{equation*}
This transformation is sufficient at one- and two-loops, but not at three-loops, where the new letters $\{\log(u^2-4v),\log(v^2-4v)\}$ first appear. 
Therefore we introduce a further transformation. The corresponding path $\gamma_\epsilon$ in the $\delta$-$t$-plane is shown in Fig.~\ref{fig:high_energy_path}(b) and it is divided into five straight sections $\gamma_\epsilon^i$, $i=1,2,...,5$. The crucial point is now that after the transformation the ambiguities are resolved and we can take the limit $\epsilon \rightarrow 0$ on the separate sections. Then we solve the differential equation in this limit. 
For example, the first segment can be parameterized by
\begin{equation*}
 \gamma_\epsilon^1(\tau)=(\delta,t)=\left(\frac{4 \epsilon}{(1-\epsilon)^2},\epsilon \tau \right) \, ,\quad \tau \in [0,1] \, \quad \epsilon>0 \, .
\end{equation*}
In the limit $\epsilon \rightarrow 0$ the alphabet becomes $\{\log(\tau)$, $\log(1 - \tau)$, $\log(1 + \tau)\}$ and therefore the solution of the differential equation is given in terms of harmonic polylogarithms. However a new subtleness arises, which we so far have not encountered. The solution exhibit logarithmic divergences at $\tau=0$ and $\tau=1$. This means we have to fix the integration constants using the asymptotic expansion at $\tau=0$. The boundary value at $\tau=0$ in the limit $\epsilon \rightarrow 0$ is $\mi{g}_{BV}^u$ from before. The differential equation on the other sections can also be solved in terms of harmonic polylogarithms in the same way. After extracting the boundary value at the end point of the last section $\gamma_\epsilon^5$ we precisely get $\mi{g}_{BV}^v$, which is a strong crosscheck for our calculations. Additionally the solution on the third section $\gamma_\epsilon^3 \big|_{\epsilon \rightarrow 0}=(\delta,t)=(0,\tau)$, $\tau \in [0,1]$ is our desired boundary value for the high energy expansion. The  boundary value depends only on the ratio $R=u/v$ via $\tau=t=1/(1+R)$. With the known identities between harmonic polylogarithms the boundary value can be rewritten such that only harmonic polylogarithms with argument $R$ appear.

Now we are finally in a position to calculate the high energy expansion. For this we transform from $(u,v)$ to $(\epsilon,R)$ using 
\begin{equation*}
 \frac{4 \epsilon}{(1-\epsilon)^2}=u+v \, , \qquad R=\frac{u}{v}
\end{equation*}
and solve the differential equation in an asymptotic expansion in $\epsilon$. 
The final result is then rewritten in $\rho=\epsilon/(1-\epsilon)^2=(u+v)/4$.

\subsection{Threshold expansion}

We now consider the threshold expansion of the amplitude. For this we solve the differential equation in an asymptotic expansion in $\beta_u=\sqrt{1+u}$ around $\beta_u=0$ in the physical region $-s<t<0$. As in the previous limits the transportation of boundary value is the most complicated part of the calculation. However we can use the results from the forward limit, where the master integrals were calculated in the limit $v \rightarrow \infty$ for arbitrary $4m^2<s$. As the next step we extract the boundary value at the threshold $\beta_u=0$. So far we only have considered the case $v \rightarrow \infty$ or equivalently $t \rightarrow 0^-$, but we are interested in an expansion in $\beta_u$ for arbitrary $-s<t<0$ or equivalently $1<v<\infty$. Therefore we solve the differential equation at the threshold in the new variable
\begin{equation} \label{Trafo_Threshold}
 v=\frac{4(1-y)^2}{(1-(1-y)^2)^2} \, , \qquad \text{with:} \; y \in(0,2-\sqrt{2})
\end{equation}
in terms of iterated integrals. This variable transformation is chosen such that the alphabet does not contain any square roots; it is polynomial, with the highest degree being four. For brevity we do not write it down here. The integration constants of the solution of the differential equation are fixed by the previously extracted boundary value. Note that the solution does not have any divergences at $y=0$ or equivalently $v=\infty$. 
 Finally we us this solution as the new boundary value for the asymptotic expansion of the master integrals in $\beta_u$ around $\beta_u=0$. We are only interested in the imaginary part of the amplitude, which simplifies the result significantly. The imaginary part of the amplitude depends only polynomially on $t$, whereas the real part contains iterated integrals.

\section{Soft current computation}  
\label{app_soft_current}

For the one- and two-loop calculation we used the Feynman diagrammatic approach and evaluated the correlators in momentum space. The diagrams where generated with QGRAF \cite{Nogueira:1991ex} and the output further processed with a custom Mathematica code which expresses the result in terms of a number of scalar loop integrals. 
To treat the Majorana fermions we used the techniques described in \cite{Dreiner:2008tw}. 
The integral reduction to a set of master integrals was done with FIRE5 \cite{Smirnov:2008iw, Smirnov:2013dia, Smirnov:2014hma} in combination with LiteRed \cite{Lee:2012cn, Lee:2013mka}. As a non trivial cross-check the calculation was done in covariant gauge. 
The gauge parameter drops out after the reduction to master integrals. In addition our Mathematica code was tested by reproducing the two-loop cusp anomalous dimension \cite{Correa:2012nk,Henn:2013wfa,Grozin:2015kna} and the two-loop jet function in soft-collinear effective theory \cite{Becher:2006qw}. 

In this appendix we discuss the calculation of the one- and two-loop master integrals with the differential equation method \cite{Kotikov:1990kg, Kotikov:1991pm, Bern:1993kr, Remiddi:1997ny, Gehrmann:1999as, Henn:2013pwa,Henn:2014qga}. We explain the one-loop case in detail. As a cross-check we compared the analytic results for the master integrals on several kinematic points in the Euclidean region with the numerical results obtained from FIESTA4 \cite{Smirnov:2015mct}. We found perfect agreement within the error bars.

\subsection{One-loop master integrals}

\begin{figure}[t]
\centering

     \centering	  
          \begin{tikzpicture}[scale=2.2]

	  \coordinate (cusp) at (0,0) {};
	  \coordinate (InW) at (-1,-1) {};
	  \coordinate (OutW) at (1,-1) {};
	  \coordinate (InPhi) at (0,-1) {};
	  \coordinate (le) at (-0.7,-0.7) {};
	  \coordinate (ri) at (+0.7,-0.7) {};
	  \coordinate (v1) at (0,-0.7) {};
	  \coordinate (v2) at (+0.35,-0.35) {};
	  
	  \draw[Wilson_blank] (le) -- node[anchor=south east, midway] {$a_1$} (cusp)  -- node[anchor=south west, midway] {$a_2$} (ri) {};
	  \draw[Wilson_arrow] (InW)  --   (le)  node[pos=0.3,anchor=west] {$\; v_1$};
	  \draw[Wilson_arrow] (ri) --  (OutW)  node[pos=0.7,anchor=east] {$v_2 \;$};
	  \draw[style={postaction={decorate}, decoration={markings,mark=at position .6 with {\arrow{Stealth[scale=1.1]}}}}] (InPhi) --  (v1) node[pos=0.3,anchor=west] {$p_3$};
	  \draw (ri) -- node[anchor=south, pos=0.6] {$a_4$} (v1) -- node[anchor=south, pos=0.4] {$a_3$} (le) {};


      \end{tikzpicture}

      
\caption{One-loop soft current integral topology}
\label{fig:One_Loop_Topo}
\end{figure}
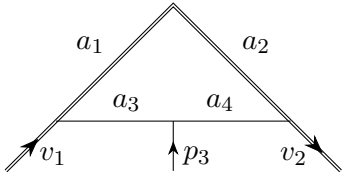

Let us briefly recall the kinematics of our problem. With the two directions of the Wilson lines $v_1$ and $v_2$, and the on-shell momentum $p_3$ of the external scalar, we can build three invariants
\begin{align}
 & \cos(\phi) = v_1 \cdot v_2 = \frac{1}{2} \left(x+\frac{1}{x} \right)\,, & & s_1=-2v_1 \cdot p_3 & & s_2=+2v_2 \cdot p_3 \, ,
\end{align}
where we used $v_1^2=v_2^2=1$ and $p_3^2=0$. 
In the following, the variable  $x=e^{i \phi}$ turns out to be most useful.
At this point it is also worthwhile to introduce the Gram determinant $G(q_1,q_2,q_3) = \det( q_{i}\cdot q_{j})$  formed by the three vectors. 
It is given by
 \begin{equation} \label{Gramm_Det}
  G(v_1,v_2,p_3) =-
  \frac{1}{4 x}(s_1 x+s_2)(s_2 x+ s_1) \, .
 \end{equation}
 If all the vectors lie in the same plane the Gram determinant vanishes. 
 This hypersurface will turn out to be useful later when determining boundary values for the differential equations.

The one loop integral family is defined as 
\begin{align}
G_{a_1,a_2,a_3,a_4}  = e^{\epsilon \,  \gamma_{E}}  \int \frac{\d^{D}k}{i \pi^{D/2}} \frac{1}{D_1^{a_1} \, D_2^{a_2} \,D_3^{a_3} \,D_4^{a_4}} \, ,
\end{align}
where $a_k \in \mathbb{Z}$ and the propagators are given by
\begin{align}
 & D_1=-2 (k+p_3) \cdot v_1 \, , & 	& D_3=-(k+p_3)^2  \,, \nonumber \\
 & D_2=-2 k\cdot v_2 \, , &	& D_4=-k^2 \, .
\end{align}
See Fig.~\ref{fig:One_Loop_Topo}.
The integral family has five master integrals. In the one loop case it is possible to find a basis, where all master integrals have uniformal transcendental weight. Such a basis is called UT or canonical basis. The master integrals $\vec{g}$ of an UT basis fulfill a particular nice differential equation \cite{Henn:2013pwa}
\begin{align} \label{DE_UT}
\d \vec{g}(x,s_1,s_2) = \epsilon \, \d \tilde{A}(x,s_1,s_2) \, \vec{g}(x,s_1,s_2) \, ,
\end{align}
where the $\epsilon$ dependence is completely factorized and the matrix $\tilde{A}$ is a linear combination of logarithms with coefficients given by rational matrices. The set of all different logarithm appearing in $\tilde{A}$ is called alphabet of the differential equation. A possible choice of a UT basis is given by \footnote{We mark the one-loop master integrals with a tilde to distinguish them from the two-loop master integrals.}
\begin{align}
 &\tilde{g}_1= \frac{1}{2}\epsilon s_2 G_{0,1,2,0} & &\tilde{g}_2=-\frac{1}{2} \epsilon s_1 G_{1,0,0,2}  & &\tilde{g}_3= \epsilon^2  \frac{1-x^2}{x} G_{1,1,0,1} \\
 &\tilde{g}_4= \epsilon^2 \frac{1-x^2}{x} G_{1,1,1,0} & &\tilde{g}_5= -\frac{1}{4} \epsilon^2 s_1 s_2 G_{1,1,1,1} \, .  & & 
\end{align}
The alphabet consists of seven letters $\{\log(1+x)$, $\log(x)$, $\log(1-x)$, $\log(s_1)$, $\log(s_2)$, $\log(s_1 x+s_2)$, $\log(s_2 x+s_1)\}$. 
We remark that the last two letters appear as factors in the Gram determinant (\ref{Gramm_Det}).

We solve the differential equation in the variable $x$ for arbitrary $s_1$ and $s_2$ in an $\epsilon$-expansion $\vec{g}=\sum_{k=1}\epsilon^k \vec{g}^{(k)}$ using iterated integrals. 
Then the integration constants are functions of $s_1$ and $s_2$.
It turns out that  
they are determined by the analytic properties of the integrals in certain limits of $x$, 
and in terms of the two (trivial) bubble integrals
 \begin{align}
  &\tilde{g}_1= - \, \frac{1}{2}e^{\epsilon \,  \gamma_{E}} \Gamma(1+2 \epsilon) \Gamma(1-\epsilon) \, (s_2)^{-2 \epsilon} \,,  \\
  &\tilde{g}_2= \frac{1}{2}e^{\epsilon \,  \gamma_{E}} \Gamma(1+2 \epsilon) \Gamma(1-\epsilon) \, (s_1)^{-2 \epsilon} \, .
 \end{align}
In this way we get the full solution of the differential equation. 
The feature that boundary values can be obtained trivially from the differential equations and physical considerations appears to be rather general, and has been observed in many calculations, see e.g. \cite{Henn:2013nsa}.

In order to understand how to obtain the boundary values, we only need to consider two limits, as will be described presently.
First we consider the limit of a straight Wilson line  $(v_1=v_2)$, where we have $x=1$. Physically we do not expect any divergent behavior near $x=1$, but the letter $\log(1-x)$ can in principle give rise to logarithmic divergences. Such divergences can be extracted from the iterated integrals using the shuffle algebra they fulfill. The condition that there are no such divergences then yields a linear system of equations. 
The second limit is approached when all external vectors lie in the same plane. 
Then the Gram determinant (\ref{Gramm_Det}) vanishes, hence one factor $(s_1 x+s_2)$ or $(s_2 x+ s_1)$ must vanish. We expect the integrals to remain finite in this limit.
Assuming $s_2>s_1>0$, it is sufficient to consider the limit $x  \rightarrow -s_1/s_2$. The calculation is then identical to the first limit, except that we leave the Euclidean region to analytically continue the iterated integrals. For this we extract the logarithms $\log(x)$ from the iterated integrals using the shuffle algebra and use $\log(-x+i0^+)=\log(x)+i \pi$.

We need the one-loop master integrals to order $\mathcal{O}(\epsilon^3)$ for the two loop renormalization. To this order all master integrals can be expressed in terms of harmonic polylogarithms with the arguments $x$, $s_1/s_2$, $x s_1/s_2$ and $x s_2/s_1$.

\subsection{Two-loop master integrals}

We can express all planar Feynman diagrams needed in our calculation as subdiagrams of a single integral family. 
This is particularly straightforward to see when using dual coordinates.
We define the integral family as
\begin{align}
G_{a_1,\dots,a_9}  = e^{2 \epsilon \,  \gamma_{E}}  \int \frac{\d^{D}k_1}{i \pi^{D/2}} \int \frac{\d^{D}k_2}{i \pi^{D/2}} \prod_{k=1}^9 \, \frac{1}{D_k^{a_k}}\, ,
\end{align}
where $a_k \in \mathbb{Z}$ and the propagators are given by
\begin{align}
 & D_1=-2 k_1 \cdot v_1 \,, 		&	& D_4=-k_1^2 \,,		&	& D_7=-k_2^2 \,,	\nonumber	\\
 & D_2=-2 (k_1-p_3) \cdot v_2  \,, 	&	& D_5=-2(k_2+p_3) \cdot v_1 \,, & 	& D_8=-(k_2+p_3)^2 \,,	\nonumber	\\
 & D_3=-(k_1-p_3)^2 \,,			&	& D_6=-2 k_2 \cdot v_2  \,, 	&	& D_9=-(k_1-k_2-p_3)^2 \,.
\end{align}

\begin{figure}[t]
\centering

 \begin{minipage}{.48\textwidth} 
     \centering

        \begin{tikzpicture}[scale=2.7]

	  \coordinate (cusp) at (0,0) {};
	  \coordinate (InW) at (-1,-1) {};
	  \coordinate (OutW) at (1,-1) {};
	  \coordinate (InPhi1) at (-0.25,-1) {};
	  \coordinate (InPhi2) at (0.25,-1) {};
	  \coordinate (le1) at (-0.75,-0.75) {};
	  \coordinate (le2) at (-0.25,-0.25) {};
	  \coordinate (ri1) at (+0.75,-0.75) {};
	  \coordinate (ri2) at (+0.25,-0.25) {};
	  \coordinate (v1) at (-0.25,-0.75) {};
	  \coordinate (v2) at (+0.25,-0.75) {};
	  
	  \draw[Wilson_blank] (le1) -- node[pos=0.5,anchor=south east] {$a_1$} (cusp)  -- node[pos=0.5,anchor=south west] {$a_2$} (ri2) --  node[pos=0.5,anchor=south west] {$a_6$} (ri1) {};
	  \draw[Wilson_arrow] (InW)  --   (le1)  node[pos=0.3,anchor=west] {$\; v_1$};
	  \draw[Wilson_arrow] (ri1) --  (OutW) node[pos=0.7,anchor=east] {$v_2 \;$};
	  \draw[style={postaction={decorate}, decoration={markings,mark=at position .6 with {\arrow{Stealth[scale=1.1]}}}}] (InPhi2) --  (v2) node[pos=0.3,anchor=east] {$p_3$};
	  \draw (ri1) -- node[pos=0.55,anchor=south] {$a_7$} (v2) -- node[pos=0.5,anchor=south] {$a_8$} (v1) -- node[pos=0.45,anchor=south] {$a_4$} (le1) {};
	  \draw (v1) -- node[pos=0.5,anchor= south east] {$a_9$} (ri2) {};

	  \node at (-1,0) {(a)};
	  

      \end{tikzpicture}
 \end{minipage}     
  \hfill  
 \begin{minipage}{.48\textwidth} 
     \centering

        \begin{tikzpicture}[scale=2.7]

	  \coordinate (cusp) at (0,0) {};
	  \coordinate (InW) at (-1,-1) {};
	  \coordinate (OutW) at (1,-1) {};
	  \coordinate (InPhi1) at (-0.25,-1) {};
	  \coordinate (InPhi2) at (0.25,-1) {};
	  \coordinate (le1) at (-0.75,-0.75) {};
	  \coordinate (le2) at (-0.25,-0.25) {};
	  \coordinate (ri1) at (+0.75,-0.75) {};
	  \coordinate (ri2) at (+0.25,-0.25) {};
	  \coordinate (v1) at (-0.25,-0.75) {};
	  \coordinate (v2) at (+0.25,-0.75) {};
	  
	  \draw[Wilson_blank] (le1) -- node[pos=0.5,anchor=south east] {$a_1$} (le2)  -- node[pos=0.5,anchor=south east] {$a_5$} (cusp) --  node[pos=0.5,anchor=south west] {$a_6$} (ri1) {};
	  \draw[Wilson_arrow] (InW)  --   (le1)  node[pos=0.3,anchor=west] {$\; v_1$};
	  \draw[Wilson_arrow] (ri1) --  (OutW) node[pos=0.7,anchor=east] {$v_2 \;$};
	  \draw[style={postaction={decorate}, decoration={markings,mark=at position .6 with {\arrow{Stealth[scale=1.1]}}}}] (InPhi1) --  (v1) node[pos=0.3,anchor=west] {$p_3$};
	  \draw (ri1) -- node[pos=0.55,anchor=south] {$a_7$} (v2) -- node[pos=0.5,anchor=south] {$a_3$} (v1) -- node[pos=0.45,anchor=south] {$a_4$} (le1) {};
	  \draw (v2) -- node[pos=0.6,anchor=south west] {$a_9$} (le2) {};

	  \node at (-1,0) {(b)};
	  

      \end{tikzpicture}
 \end{minipage}            
      
      
\caption{Two-loop soft current integral topologies.}
\label{fig:Topos}
\end{figure}
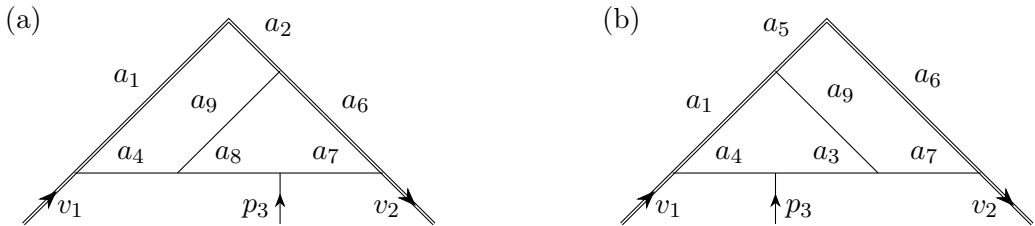

The subtopologies we need for the calculation of the master integrals appearing in the correlators are defined by $G_{a_1,a_2,b_3,a_4,b_5,a_6,a_7,a_8,a_9}$ and $G_{a_1,a_2,a_3,b_4,a_5,b_6,a_7,a_8,a_9}$, where $a_k$ is an arbitrary integer and $b_k$ is zero or a positive integer.
They are related to each other through interchanging $v_2 \leftrightarrow -v_1$. In Fig.~\ref{fig:Topos} both subtopologies are shown. Note that in the Feynman diagrams more subtopologies appear, but after integral reduction all master integrals can be mapped onto these two.
For these subtopologies we have in total 47 master integrals. 

As in the one-loop case, we derived differential equations for all of them.
Fixing the boundary values is analogous to the one loop case. 
In order to fix all boundary values we computed the following integrals
\begin{align}
 g_1 &= \frac{1}{2} \epsilon^2 s_2 \, G_{0, 0, 0, 2, 0, 1, 0, 0, 2}= \frac{1}{2} e^{2 \epsilon \,  \gamma_{E}} \Gamma^2(1-\epsilon) \Gamma(1+4 \epsilon) (s_2)^{-4 \epsilon} \\
 g_6 &= \frac{1}{4} \epsilon^2 s_2^2 \,G_{0, 1, 0, 2, 0, 1, 0, 2, 0}= \frac{1}{4} e^{2 \epsilon \,  \gamma_{E}} \Gamma^2(1-\epsilon) \Gamma^2(1+2 \epsilon) (s_2)^{-4 \epsilon} \\
 g_{16}& = \frac{1}{2} \epsilon^3 s_2 \,  G_{0, 2, 0, 1, 0, 1, 1, 0, 1}= \left[-\frac{\pi ^2 \epsilon^2}{48}+ \frac{9 \epsilon^3 \zeta_3}{8}+\mathcal{O}\left(\epsilon^4 \right) \right] (s_2)^{-4 \epsilon}
\end{align}
and their symmetric counterparts with $v_2 \rightarrow -v_1$.
The last integral was calculated with Mellin-Barnes techniques using the Mathematica package MB.m \cite{Czakon:2005rk}.

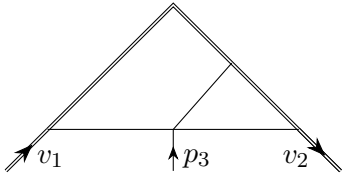
\begin{figure}[t]
\centering

     \centering	  
          \begin{tikzpicture}[scale=2.2]

	  \coordinate (cusp) at (0,0) {};
	  \coordinate (InW) at (-1,-1) {};
	  \coordinate (OutW) at (1,-1) {};
	  \coordinate (InPhi) at (0,-1) {};
	  \coordinate (le) at (-0.75,-0.75) {};
	  \coordinate (ri) at (+0.75,-0.75) {};
	  \coordinate (v1) at (0,-0.75) {};
	  \coordinate (v2) at (+0.35,-0.35) {};
	  
	  \draw[Wilson_blank] (le) -- (cusp)  -- (ri) --  (OutW) {};
	  \draw[Wilson_arrow] (InW)  --   (le)  node[pos=0.3,anchor=west] {$\; v_1$};
	  \draw[Wilson_arrow] (ri) --  (OutW)  node[pos=0.7,anchor=east] {$v_2 \;$};
	  \draw[style={postaction={decorate}, decoration={markings,mark=at position .6 with {\arrow{Stealth[scale=1.1]}}}}] (InPhi) --  (v1) node[pos=0.3,anchor=west] {$p_3$};
	  \draw (v1) -- (v2) {};
	  \draw (ri) -- (v1) -- (le) {};


      \end{tikzpicture}

      
\caption{Elliptic sector}
\label{fig:Elliptic}
\end{figure}

An interesting new feature of this calculation is that there is a sector of the differential equations that leads to elliptic polylogarithms. 
The relevant integral sector is shown in Fig.~\ref{fig:Elliptic}. The elliptic nature of the integral can be seen by considering the projection of the differential equations onto that sector, as we describe below. The latter contains two master integrals. 

Our expectation was that the elliptic sector should be irrelevant for the computation of the divergent part of the correlator (\ref{Cor_Cusp}). Given this expectation, we made a choice of master integrals that have good infrared and ultraviolet properties, namely $g_{38}  =\epsilon^4 G_{1, 1, 0, 1, 0, 1, 1, 0, 1}$ and $g_{39}  =\epsilon^4 G_{1, 1, -1, 1, 0, 1, 1, 0, 1}$.
Indeed, with this choice it turns out that the elliptic sector decouples from the integrals needed for our correlator, to the order in $\epsilon$ that was required.
The remaining 37 master integrals with fewer propagators than the elliptic sector are in UT form (\ref{DE_UT}), with the same alphabet as the one-loop master integrals. 
As in the one-loop case, up to the order in $\epsilon$ that is needed for the renormalization of the correlators (\ref{Renormalization_Condition}), all master integrals at two loops can be expressed in terms of harmonic polylogarithms.

For this reason a further analysis of the elliptic sector was not needed here. However, we do wish to make a few observations that may be of interest for future studies.
The differential equation w.r.t. $x$, projected onto the elliptic sector, i.e. neglecting contribution from the lower sectors, is given by
\begin{align}
\frac{\partial}{\partial x}   \left(
\begin{array}{c}
g_{38} \\
g_{39} 
\end{array}
\right)
_{\text{max cut}}  = {A}_{\mathrm{elliptic}}  \,  \left(
\begin{array}{c}
g_{38} \\
g_{39} 
\end{array}
\right)
_{\text{max cut}}  \,,
\end{align}
with
\begin{align}
{A}_{\mathrm{elliptic}} =\left(
\begin{array}{cc}
 \frac{(6 \epsilon -1) s_1 s_2 (x-1) (x+1)}{4 x \left(x s_1+s_2\right) \left(s_1+x s_2\right)} & \frac{-(6 \epsilon -1) (x-1) (x+1) \left(s_2 x^2+2 s_1 x+s_2\right)}{4 x^2 s_2 \left(x s_1+s_2\right)
   \left(s_1+x s_2\right)} \\
 \frac{(2 \epsilon -1) s_1^2 s_2 \left(s_2 x^2+2 s_1 x+s_2\right)}{4 (x-1) (x+1) \left(x s_1+s_2\right) \left(s_1+x s_2\right)} & \frac{(2 \epsilon -1) \left[8 x \left(x^2+1\right) \left(s_1^2+s_2^2 \right)+\left(7
   x^4+18 x^2+7\right) s_2 s_1\right]}{4 x (x-1)  (x+1) \left(x s_1+s_2\right) \left(s_1+x s_2\right)} \\
\end{array}
\right)
\end{align}
Note that the matrix is affine in $\epsilon$.  One could  simplify it further, but this will not be discussed here. See ref. \cite{Henn:2014qga} and references therein for a general discussion.

In order to bring this differential equation into canonical form, one needs to solve the differential equation at $\eps=0$.
It is sufficient to find the solution for one of the two solutions, as the other one can be obtained via the coupled equations.
Here we focus on the  integral $ g_{38}$. This integral is both UV and IR finite.

The desired solution at $\eps=0$ can be found from the maximal cut of the integral \cite{Bonciani:2016qxi,Primo:2016ebd}, which is the natural generalization of the leading singularities \cite{Henn:2013pwa} to the elliptic case. Using the Baikov representation \cite{Baikov:1996iu} and a loop by loop approach for calculating the maximal cut \cite{Frellesvig:2017aai,Harley:2017qut}, we get the homogeneous solutions 
\begin{align}
 & \frac{1}{\sqrt{a_2}} \mathrm{K}\left( \frac{a_1}{a_2} \right) \, , & & \frac{1}{\sqrt{a_2}} \mathrm{K}\left(1- \frac{a_1}{a_2} \right) \, ,
\end{align}
where $\mathrm{K}$ is the complete elliptic integral of the first kind and we have
\begin{equation}
 a_{1/2}= \frac{s_2}{2} \left[s_1 \left(x+\frac{1}{x}\right)+ 2 s_2 \mp 2 \sqrt{\frac{1}{x} \left(s_1 x+s_2\right) \left(s_2 x+s_1\right)}\right] \, .
\end{equation}

\bibliographystyle{JHEP}

\bibliography{bibfile}

\end{document}